\documentclass[aps,prl,superscriptaddress,showpacs,floatfix,column]{revtex4-1}
\usepackage{graphicx}	
\usepackage{dcolumn}	
\usepackage{xspace}	
\usepackage{multirow}

\newcommand{\snn}{\ensuremath{\sqrt{s_{\rm NN}}}}
\newcommand{\pT}{\ensuremath{p_{T}}}
\newcommand{\ee}{\ensuremath{e^{+}+e^{-}}}
\newcommand{\pp}{\ensuremath{pp}}
\newcommand{\nch}{\ensuremath{N_{\rm ch}}}
\newcommand{\npart}{\ensuremath{N_{\rm part}}}
\newcommand{\ncoll}{\ensuremath{N_{\rm coll}}}
\newcommand{\nmono}{\ensuremath{N_{\rm mono}}}
\newcommand{\nmulti}{\ensuremath{N_{\rm multi}}}
\newcommand{\ymono}{\ensuremath{Y_{\rm mono}}}
\newcommand{\ymulti}{\ensuremath{Y_{\rm multi}}}

\newcommand{\pythia}{{\sc Pythia}}
\newcommand{\hijing}{{\sc Hijing}}
\newcommand{\raa}{\ensuremath{R_{AA}}}
\newcommand{\raanpart}{\ensuremath{\raa^{\npart}}}

\begin{document}

\title{Participant-like scaling behavior of multiplicity and charged hadron spectra in relativistic heavy ion collisions.}

\author{R. S. Hollis}
\author{A. Iordanova}
\affiliation{University of California, Riverside, CA, 92521}
\author{D. J. Hofman}
\affiliation{University of Illinois at Chicago, Chicago, IL, 60607}
\date{\today}

\begin{abstract}
We present a method for parameterizing the charged particle
multiplicity and charged hadron spectra from heavy ion data
with a simple, Glauber-inspired, model.  The basis of this
model is derived from the observation of leading hadrons in
\pp~collisions and the number of interactions calculated by
a Glauber model.  Singly hit and multiply hit nucleons are
treated as different sub-components of the same collision.
With this scheme, for fixed sub-component yields, we find that the multiplicity and charged
hadron spectra can be reproduced, without the need for large
suppression at high-\pT.  Suppression is still observed, in
the low- to intermediate-\pT~region, but this is confined to
suppression of surface emission partons.  At high-\pT, the
suppression disappears entirely, leaving only a possible 
Cronin-type enhancement.  The suppression observed in 200\,GeV
collisions is found to be the same for 62.4\,GeV data.
\end{abstract}

\maketitle

\section{Introduction}
The comparison of hadronic collider (\pp~collisions) and electron
collider (\ee~collisions) data have shown a clear discrepancy in
the produced particle multiplicity for collisions measured at the
same center-of-mass
energy.  This has long been explained by the observation that
hadronic interactions can be decomposed in terms of ``collision
products'' and ``leading hadrons'' -- with the latter carrying away
(on average) half of the total energy available for particle
production~\cite{cite:LeadingHadrons}.  It is found that one can
unify these data by shifting the center-of-mass energy of the
\pp~collisions to an effective center-of-mass energy, which is half
of the original collision energy, empirically determined.
Superimposing central heavy ion data onto this agrees
more closely with the \ee~data than the hadronic collision (\pp)
data~\cite{cite:PHOBOS_Universality}.  Combining all these
observations, one can surmise that collisions in which all
``constituents'' of the proton are used (i.e. no leading hadron) will
produce higher (\ee-like) multiplicities.

The model presented here, originally described in
Ref.~\cite{cite:RichardsThesis}, uses these ideas as a premise
to divide the
participant region in heavy ion collisions into two distinct
regions; these are noted as ``mono'' and ``multi''.  A Glauber
model~\cite{cite:Glauber} is used to calculate the number of 
participants and the number of times each nucleon is struck
by another from the opposing nucleus.  Mono refers to nucleons
which undergo a single hit -- similar to \pp~interactions, which
presumably can ``liberate'' a leading hadron.  Multi refers to
nucleons which are multiply struck (by two or more) nucleons,
however, these are only counted once and thus  do not
accumulate as quickly as the number of binary collisions.

\begin{figure*}[ht]
\centering
\includegraphics[angle=0,width=0.95\textwidth]{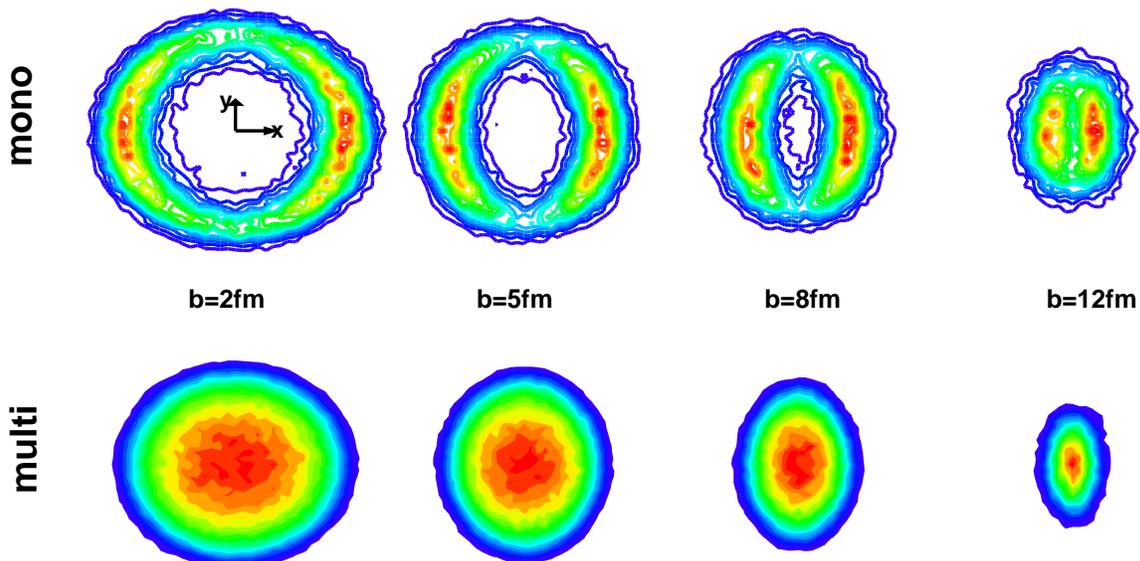}
\vspace{-20pt}
\caption{\label{fig:MonoMultiPositions}
(color online) Predicted relative positions of mono- (upper) and
multi-collisions (lower).  The distributions are based on a Glauber
calculation at impact parameter $b$\,=2,\,5,\,8,\,12\,fm
(\npart\,$\sim$\,359,\,261,\,140,\,24, \nmono\,$\sim$\,28,\,38,\,33,\,12, and
\nmulti\,$\sim$\,331,\,223,\,107,\,12).
In this picture, the reaction plane is always along the $x$-axis.}
\end{figure*}

This phenomenological approach is similar to the successful
core/corona approach of EPoS~\cite{cite:CoreCorona}.
The main technical difference surrounds the expected position
with respect to the reaction plane, of the mono (or corona)
nucleons.  The current model expects that the peak of mono
nucleons to be {\it along} the reaction plane (see
Fig.~\ref{fig:MonoMultiPositions}), whereas in the corona from EPoS
is expected to peak 90$^{0}$ away~\cite{cite:CoreCorona}.  A more
recent `toy model' implementation by the EPoS
authors~\cite{cite:CoreCorona2}, however,
does use the same ``Glauber-method'' to determine core (multi) and corona
(mono) regions to describe the strangeness enhancement observed in
Au+Au and Cu+Cu
data.  In that implementation no interactions between the core and
corona regions is assumed, each ``corona'' contributes one-half of
a minimum bias interaction.  The authors note that fixing the mono
(corona) is not necessary.  However, although they undergo different
assumptions, both can provide a good description of
the data. Where these models differ is in the extent to which
they can be (or at least have been) applied and the possible
origin of the two components, or regions, of the collision.  By using
the model presented here, it is possible to not only describe with
success low-\pT~phenomenon observed in heavy ion collisions but also
high-\pT~phenomena.  This can lead to hints as to the origin of
disappearing back-to-back jets and their reappearance at low-\pT.

\section{The Model}

\begin{figure}[t]
\centering
\includegraphics[angle=0,width=0.475\textwidth]{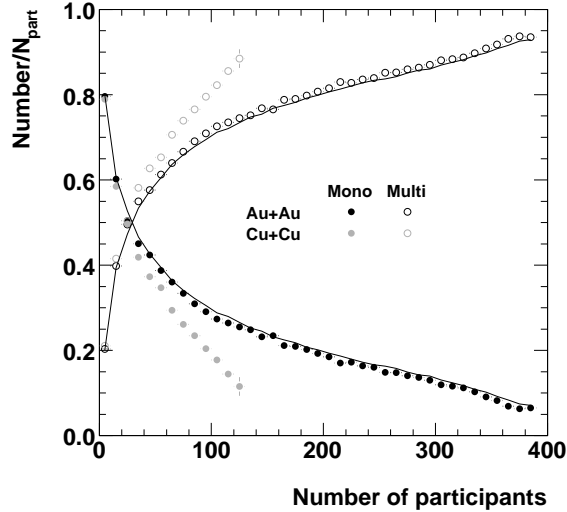}
\vspace{-20pt}
\caption{\label{fig:MonoMultiVsNpart}
The relative number of mono and multi collisions versus the number of
participants.  Black (grey) circles represent Au+Au (Cu+Cu) collisions
at 200\,GeV ($\sigma_{NN}$\,=\,42\,mb); closed and open symbols represent
mono and multi collisions respectively.  The values for 19.6\,GeV
($\sigma_{NN}$\,=\,33\,mb) Au+Au data are depicted as lines for comparison.}
\end{figure}

The implementation of this model is very simple.  A Glauber model
is used to count the number of times each nucleon is struck (the
Glauber model implementation is similar to
Ref.~\cite{cite:PHOBOS_Glauber}).  For
each Monte-Carlo ``collision'' the nucleons are sorted into three
categories: not hit (spectators), singly hit nucleons (\nmono), and
multiply hit nucleons (\nmulti).  Spectators are not considered in
this model.  After many trials, and at all possible impact parameters,
the average number of mono- and multi-assigned nucleons are
calculated.  Figure~\ref{fig:MonoMultiVsNpart} illustrates the mean
\nmono~and \nmulti~found for Au+Au (black circles) and Cu+Cu (grey)
collisions at \snn\,=\,200\,GeV.  \nmono~and \nmulti~versus centrality
are not found to appreciably change with collision energy, as shown
by lines depicting \nmono~and \nmulti~for 19.6\,GeV collisions.

After calculating the number of mono and multi participants for a 
given centrality class it is possible to form the multiplicity
(and spectra, etc.) corresponding to that data using the formula:

\begin{equation}
\label{eqn:MonoMulti}
\ensuremath{ Y_{\rm Au+Au} = \nmono Y_{\rm mono} + \nmulti Y_{\rm multi}}
\end{equation}

\noindent $Y_{\rm Au+Au}$ represents the modeled Au+Au data, $Y_{\rm mono}$
is the yield expected from singly interacting nucleons (ostensibly
minimum bias \pp~data) and $Y_{\rm multi}$ is the yield expected from
the multi underlying distribution.  $Y_{\rm mono}$ and $Y_{\rm multi}$
are fixed for a given data sample and do not change versus centrality.
$Y$ itself represents any distribution, for example multiplicity,
charged hadron spectra, kinetic freeze-out temperature, or any other
observable.

In this paper, we first explore the possible
origin of the underlying multi distribution which could be from one
of many different sources such as non-single diffractive (NSD) events
(unlikely as these may still have leading hadrons),
high-multiplicity events (for example greater than
$\langle\nch^{pp}\rangle$/2.), high-\pT~data ($\hat{p}_{T}$$>\sim$1.8
to represent the close-packedness of the multi nucleons), a
gluon-only source distribution.  We use each of these different
hypotheses to test the systematic dependence -- although not to
choose a preferential particle production mechanism but rather to
provide a proof of principle that one can reasonably reproduce the
particle distributions in heavy ion collisions from these simple sources.
We note that a single distribution for mono (\pp~minimum bias) and a
single distribution for multi is assumed, which is sufficient to
reproduce all the data versus centrality.
The centrality dependence observed in the data is then entirely due to
the interplay between \nmono~and \nmulti, Fig.~\ref{fig:MonoMultiVsNpart},
and not from a changing or modified underlying distribution.

Secondly, as a more rigorous test, fits to the data to extract the underlying mono
and multi distributions will be made.  In fact this is necessary
to analyze more exotic data than multiplicity and charged hadron
spectra.  In this way we decouple ourselves from (for example)
\pythia~\cite{cite:PYTHIA} to describe the detailed identified
particle spectra.  By using Eqn.~\ref{eqn:MonoMulti} and the
measured data to simultaneously extract the underlying
distributions, both the modification to the \pp~minimum bias
collisions at the surface (mono) as well as the true underlying
multi distribution can be studied.
As Glauber has no recourse to produce particle distributions, this
model has a purely phenomenological approach to describe the heavy
ion experimental data.

One important distinction to note for the current model versus many
other two-component models is the absence of ``collision scaling''.
Other two-component models assert that one component factorizes with
the number of participants (usually associated with soft particle
production), and the second with the number of nucleon-nucleon
interactions (\ncoll~-- representing hard collisions) within the
collision~\cite{cite:TwoComponentFit,cite:STARv2Model,cite:GyulassyWang,cite:KLN}.
In the current
discussion, we assert that the number of collisions is not the
correct scaling variable.  If, as is hypothesized, upon
collision the system becomes a hot, dense QCD matter then it may become
difficult to reliably count individual nucleon-nucleon collisions.
The currently accepted approach used to validate the number of collisions
is to measure the spectrum of direct-photons in $A$+$A$ collisions,
relative to that in \pp~interactions.  As the direct photons are not
expected to interact with the medium, it is assumed that the measured
\raa\,=\,1 validates the use of \ncoll~\cite{cite:PHENIX_dirgam},
illustrating that the number of originally produced photons are
consistent with \pp.  In more current data, however, the previously
observed \ncoll~scaling at \pT\,$\sim$\,6\,GeV/$c$ is not seen at
higher momenta~\cite{cite:PHENIX_dirgamNew}, even though punch-through
high-\pT~hadrons are seen to escape the medium~\cite{cite:STAR_dijets}.
Here, we set this \ncoll~scaling assumption aside.
The number of mono and multi collisions in the current model is
less susceptible to the assumptions needed in order to count
the number of collisions, where our {\em number}
reflects an interaction volume (multi) and simply the number of
expected nucleon-nucleon scatterings on the surface (mono).

For a low number of participants (\npart$<$25), the number of mono
collisions exceeds
that of multi, and vice-versa for a high \npart.  However, as shown
in Fig.~\ref{fig:MonoMultiVsNpart}, the number of mono does not fall to
zero even for the most central events where 
approximately~7\% of the participants undergo only a single
interaction.  An important consequence in these calculations is that
the sum of mono
and multi, by definition, is fixed to be \npart.  This restricts the
growth of multi, unlike the rapidly growing number of binary collisions,
and results in multi (and mono) being more closely associated
to a ``participant'' variable.  In calculating the mono and multi, it
is found that there is a strong dependence on the collision species,
in contrast to the number of binary collisions (\ncoll) which is
essentially the same at the same \npart.  Similarly, only a weak
dependence on the collision energy (inelastic \pp~cross-section) is
observed over the energy range of the RHIC data, whereas a strong
binary collision dependence on \ncoll~is noted.

\section{Multiplicity Distributions}

As the premise of this model uses the charged particle multiplicity to
argue for distinct mono/multi regions of the collision volume it is
natural to begin with a discussion of that experimental data.  Here,
one can simply use the measured \pp~distribution as the mono and a
second underlying distribution (for example taken from \pythia) to represent
the multi.  We combine the mono/multi distributions (and similarly
for the next section) as:

\begin{equation}
\label{eqn:multiplicity}
\ensuremath{\frac{dN_{ch}^{\rm Au+Au}}{d\eta} = 
                    \nmono\frac{dN_{ch}^{\rm mono}}{d\eta} + 
                    \nmulti\frac{dN_{ch}^{\rm multi}}{d\eta}}
\end{equation}

\noindent where \nmono~and \nmulti~distributions for Au+Au
collisions are given in Fig.~\ref{fig:MonoMultiVsNpart}.  One can
hypothesize that $dN_{ch}^{\rm mono}/d\eta$ be
$dN_{ch}^{\pp}/d\eta$ (i.e. precisely from minimum bias
\pp~interactions) or it can be a free parameter in the fit.  For simplicity
in this first case, we assume the former; later, the latter is used to
test for any modification of the underlying mono distributions.

\begin{figure*}[th]
\centering
\includegraphics[angle=0,width=0.80\textwidth]{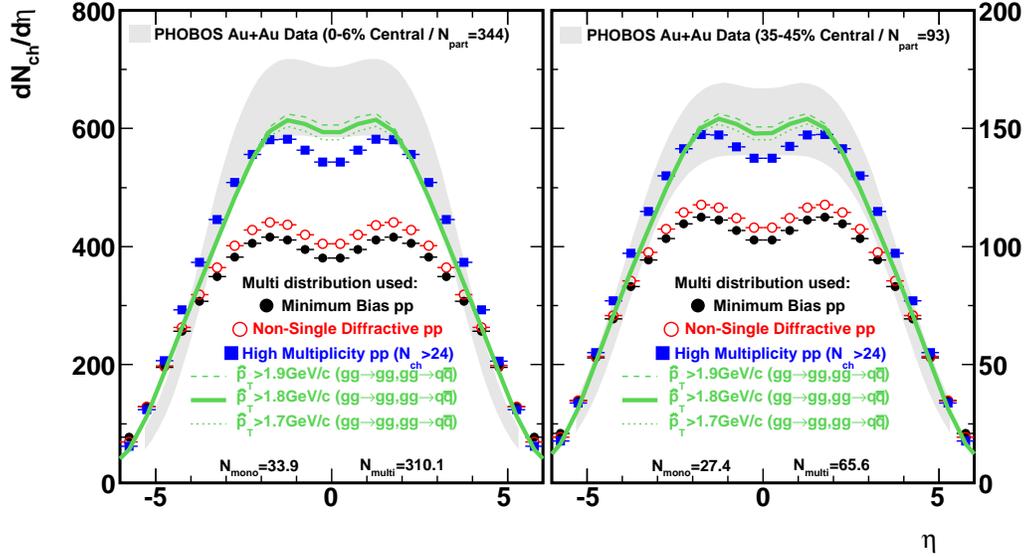}
\vspace{-24pt}
\caption{\label{fig:MultiplicityI}
(color online) Comparison between measured multiplicity versus $\eta$
(PHOBOS \snn\,=\,200\,GeV Au+Au data~\cite{cite:PHOBOS_dNdeta_200}) (band)
and model predictions with various candidates for the underlying
``multi'' $dN_{ch}^{\rm multi}/d\eta$~distribution as noted in the
legend.  0-6\% central (left) and mid-central
35-45\% (right) data are shown.  In each model representation,
the ``mono'' is fixed to the inelastic \pp~distribution at the
same energy, i.e. $dN_{ch}^{\rm mono}/d\eta$\,=\,$dN_{ch}^{\pp}/d\eta$.  Note the $y$-axis scale
difference in each figure.}
\end{figure*}

Figure~\ref{fig:MultiplicityI} shows the result of using this
formalism with several hypothesized multi multiplicity distributions
for Au+Au collisions at \snn\,=\,200\,GeV.  For the shown centrality
bins from PHOBOS~\cite{cite:PHOBOS_dNdeta_200}, one can see that the model
provides a reasonable description of the data.  Remember that the same
underlying distributions are used for both panels, only \nmono~and
\nmulti~are changed.  This is not too
surprising as the basic premise of this model is to reproduce
the centrality dependence of the charged particle multiplicity in
the context of an absence of leading hadrons in the multi region.
From this, judging the model's success is really
only dependent on the choice of the underlying multi distribution.
Three simple \pp~multiplicity-like distributions are used for the
multi component, for comparison.  The \pp~collisions at
\snn\,=\,200\,GeV minimum bias distribution is used (closed
circles) which is similar to the ``Wounded Nucleon
Model''~\cite{cite:WNM} approach and yields a multiplicity which is
$\sim$30\% lower than the measured Au+Au data.  Rejecting the single
diffractive events yields a higher multiplicity (open circles)
but is still significantly lower than the Au+Au distributions. A more
radical approach is to only choose the highest multiplicity
\pp~events, in this case more than the average ($>$24) (squares),
which does achieve the desired multiplicity.  One can
understand this approach in terms of the origins of the model --
as higher multiplicity events typically have lower energy leading
hadrons, thus releasing more energy to particle production.

Changing the reference \pp~multiplicity is not the only way to alter
the resultant multiplicity distribution.  Also shown are a subset of
events from the minimum bias \pp~cross-section which exclusively
selects out gluon interactions ($gg\rightarrow gg$ or
$gg\rightarrow q\overline{q}$) with a large momentum transfer
($\hat{p}_{T}>1.8\pm0.1$\,GeV/$c$) (lines).  (Note that the same multiplicity
enhancement is also found by using all (quark and gluon) interactions
with $\hat{p}_{T}>1.8$\,GeV/$c$.)  Such an approach may represent
the close-packedness of the nucleons which may facilitate a large
momentum transfer.  The different minimum $\hat{p}_{T}$ values
used as the multi component (denoted as dotted and dashed lines
in Fig.~\ref{fig:MultiplicityI}) show the sensitivity of the choice of
$\hat{p}_{T}$ in the final distribution
($dN_{ch}^{\rm Au+Au}/d\eta$ of Eqn.~\ref{eqn:multiplicity});
for multiplicity this is small.

\begin{figure*}[ht]
\centering
\begin{minipage}{0.23\textwidth}
\includegraphics[angle=0,width=1\textwidth]{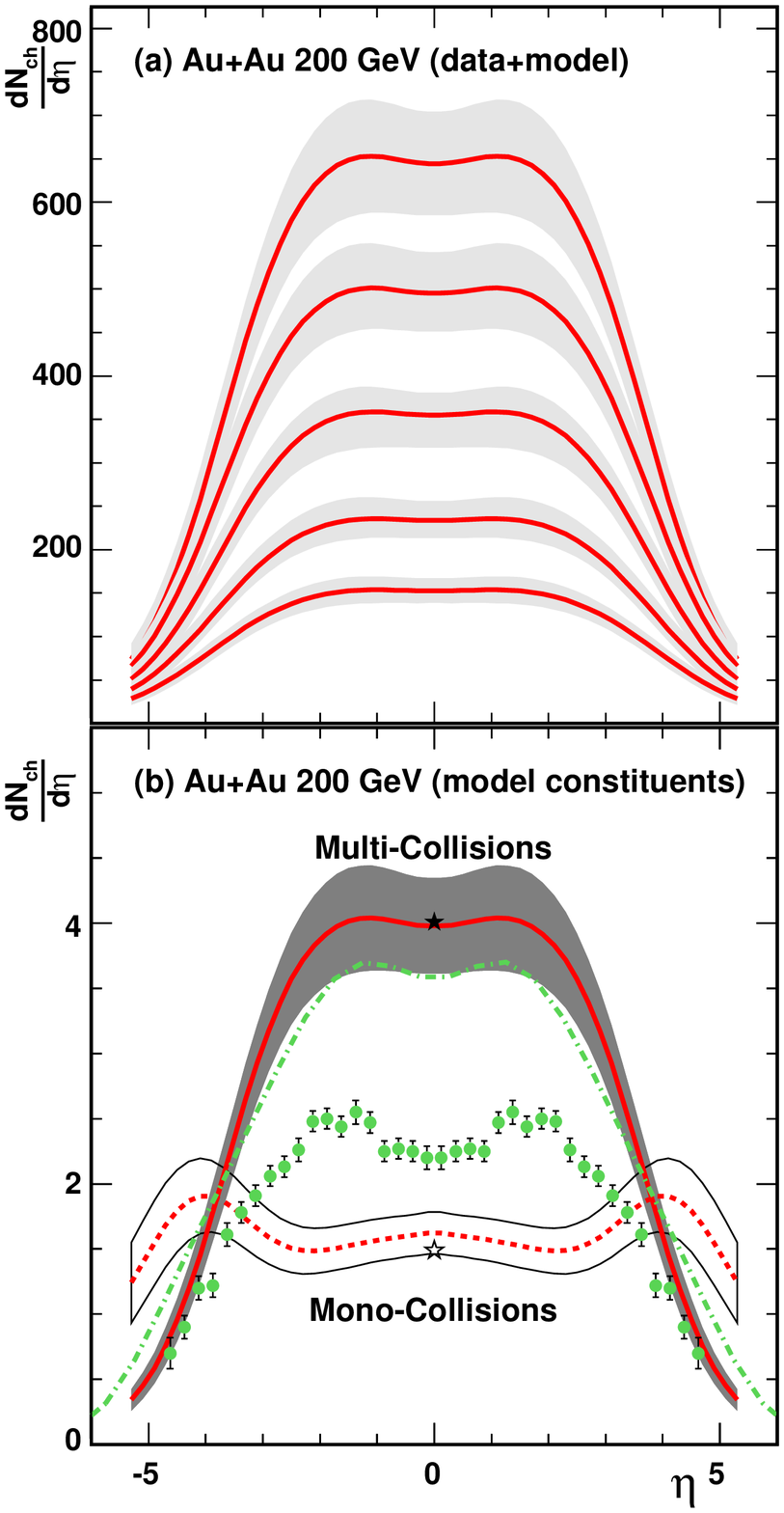}
\end{minipage}
\begin{minipage}{0.23\textwidth}
\includegraphics[angle=0,width=1\textwidth]{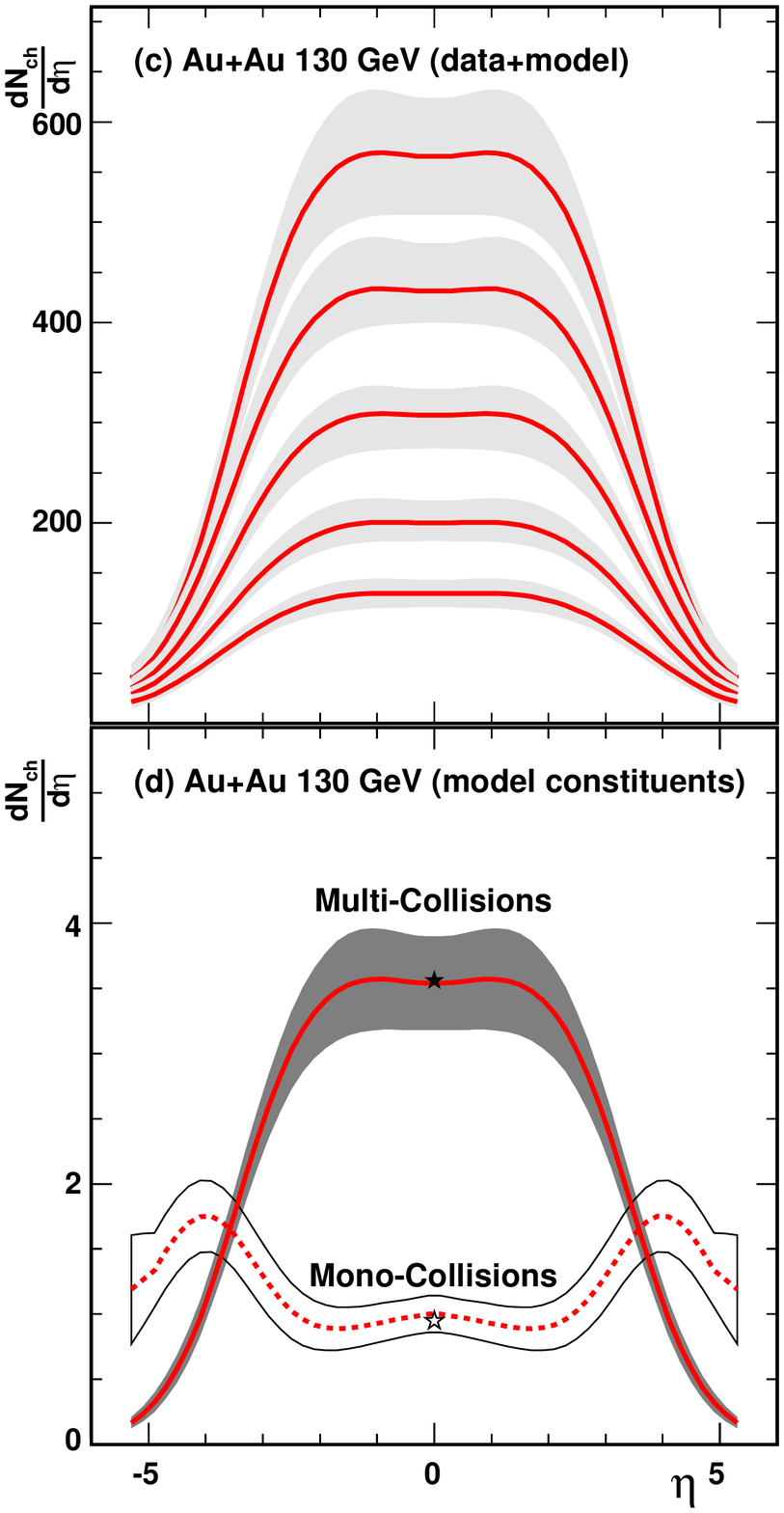}
\end{minipage}
\begin{minipage}{0.23\textwidth}
\includegraphics[angle=0,width=1\textwidth]{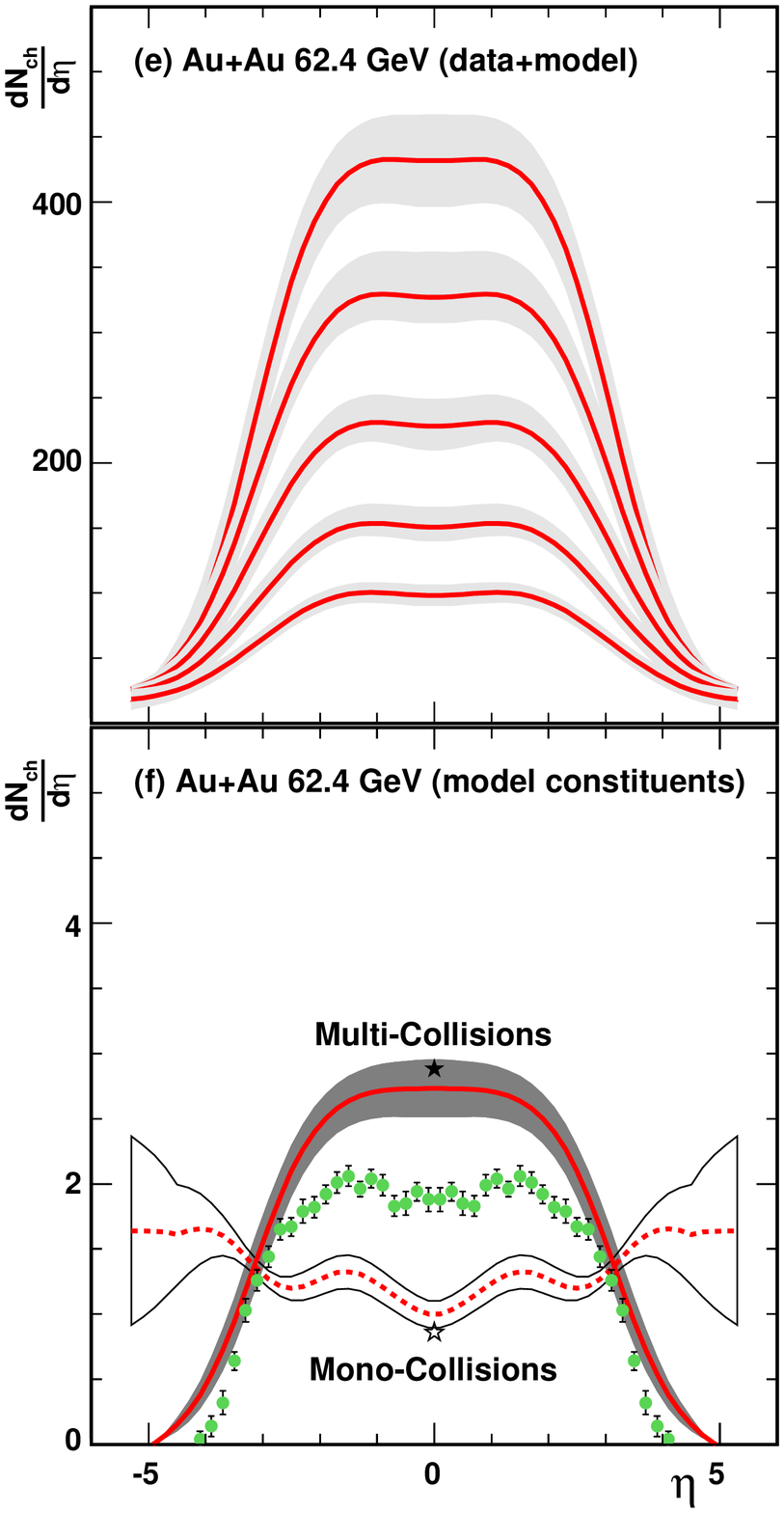}
\end{minipage}
\begin{minipage}{0.23\textwidth}
\includegraphics[angle=0,width=1\textwidth]{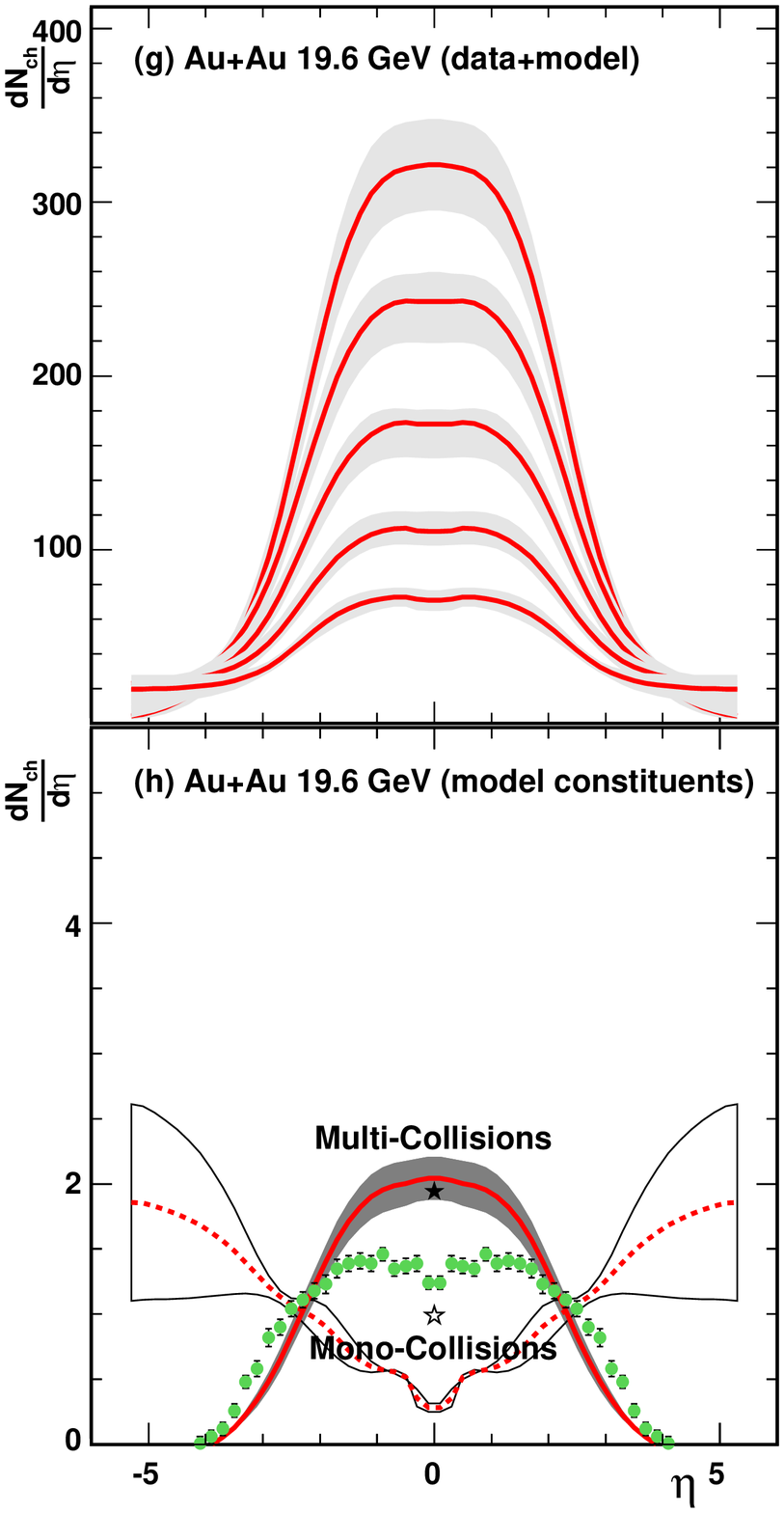}
\end{minipage}
\caption{\label{fig:MultiplicityII}
(color online) PHOBOS multiplicity data (bands) from Au+Au collisions
compared to
the multiplicity distributions derived from the model, lines, using the pair
fit method (upper panels).  Panels (a,b), (c,d), (d,e), (f,g)
show data for Au+Au collisions at 200~\cite{cite:PHOBOS_dNdeta_200}, 130~\cite{cite:PHOBOS_dNdeta_130}, 62.4~\cite{cite:PHOBOS_62.4mid}, and 19.6\,GeV~\cite{cite:PHOBOS_CuCudNdeta}
respectively.  The lower panels show the underlying \ymono~(dashed
line/open band) and \ymulti~(solid line/filled band).  For reference,
the minimum bias \pp~data~\cite{cite:ppRef200,cite:ppRef62.4_22.4}
(green closed circles) and gluon only distribution (green dot-dashed
line) are shown (the latter for 200\,GeV only).
The open (closed) star symbols represent the result from the least-$\chi^{2}$ method
to determine the underlying mono (multi) from mid-rapidity data, see
Fig.~\ref{fig:MultiplicityIImid}.
}
\end{figure*}

Rather than assuming a given \ymono~and/or \ymulti~multiplicity
distribution,
Eqn.~\ref{eqn:MonoMulti} can be used to extract these two
underlying distributions from the data.  Two bins in centrality
(preferably far apart in centrality) are used to simultaneously extract
\ymono~and \ymulti.  This is referred to as the pair
fit method.  After finding the underlying distributions, these are fixed
and all centrality bins are recombined using the \nmono~and \nmulti~as in
Eqn.~\ref{eqn:MonoMulti}.
To test the sensitivity to the bins chosen, several
combinations are made.  The top panels of Fig.~\ref{fig:MultiplicityII}
show the PHOBOS data (grey bands) with the fit distributions (solid
lines).  Panels (a,b), (c,d), (e,f), (g,h) in
Fig.~\ref{fig:MultiplicityII}, show data for Au+Au collisions at
200~\cite{cite:PHOBOS_dNdeta_200}, 130~\cite{cite:PHOBOS_dNdeta_130}, 62.4~\cite{cite:PHOBOS_62.4mid}, and 19.6\,GeV~\cite{cite:PHOBOS_CuCudNdeta} respectively.  In each case, the model
uses the most central 0-6\% and 35-45\% (mid-central) data to
extract the underlying mono and multi distributions.  Using other bin
combinations does not significantly alter the resultant distributions.
The lower panel of
Fig.~\ref{fig:MultiplicityII} shows the underlying \ymono~(dashed line) and
\ymulti~(solid line) distributions from the fit.  The grey and open bands 
represent the systematic error of the data.  For
comparison, the gluon-only distribution selected from \pythia~(dot-dashed
lines for the $\hat{p}_{T}>1.8$\,GeV/$c$ distribution only at 200\,GeV)
along with the minimum bias at the same center-of-mass energy (circles) are also shown.
The underlying multi distribution is considered ``stable'' in the sense
that using different centrality pairs does not change appreciably the extracted
\ymulti~distribution.  Mono, however, is less stable when using the pair fit method and is
found to be more dependent on the choice of centrality bin.  Several
interesting artifacts of the fit are observed.  First, the extracted
\ymulti~distribution is close to that which would be obtained
by using a gluon-only \pp~distribution in Eqn.~\ref{eqn:multiplicity},
but with more yield at mid-rapidity and also narrower.
Second, the mono distribution is considerably lower at mid-rapidity
than the measured minimum bias \pp~collisions~\cite{cite:ppRef200,cite:ppRef62.4_22.4}, as shown by the comparison of the
mono-collisions from the pair fit to minimum bias \pp~in
Fig.~\ref{fig:MultiplicityII}, panels (b), (f), and (h).
This is possibly hinting that
a modification exists, due to an opaque multi region.
A more detailed description of this modification is discussed in
the next section.  The final notable observation is that the yield
of the extracted mono interactions is higher at forward rapidities,
as compared to the minimum bias \pp.  This could possibly point to an
influence of spectators in the measured data.
To reiterate, \ymulti~represents the multiplicity per participant pair
of the multi region, which is consistent with the expected yield from
a gluon-dominated collisional center.  The core is found to be opaque,
with some of the multiplicity from the surrounding mono collisions
absorbed.  Importantly, even though the size of the central multi region
changes, the yield per participant does not change.
 
\begin{figure*}[ht]
\centering
\vspace{-11pt}
\includegraphics[angle=0,width=0.375\textwidth]{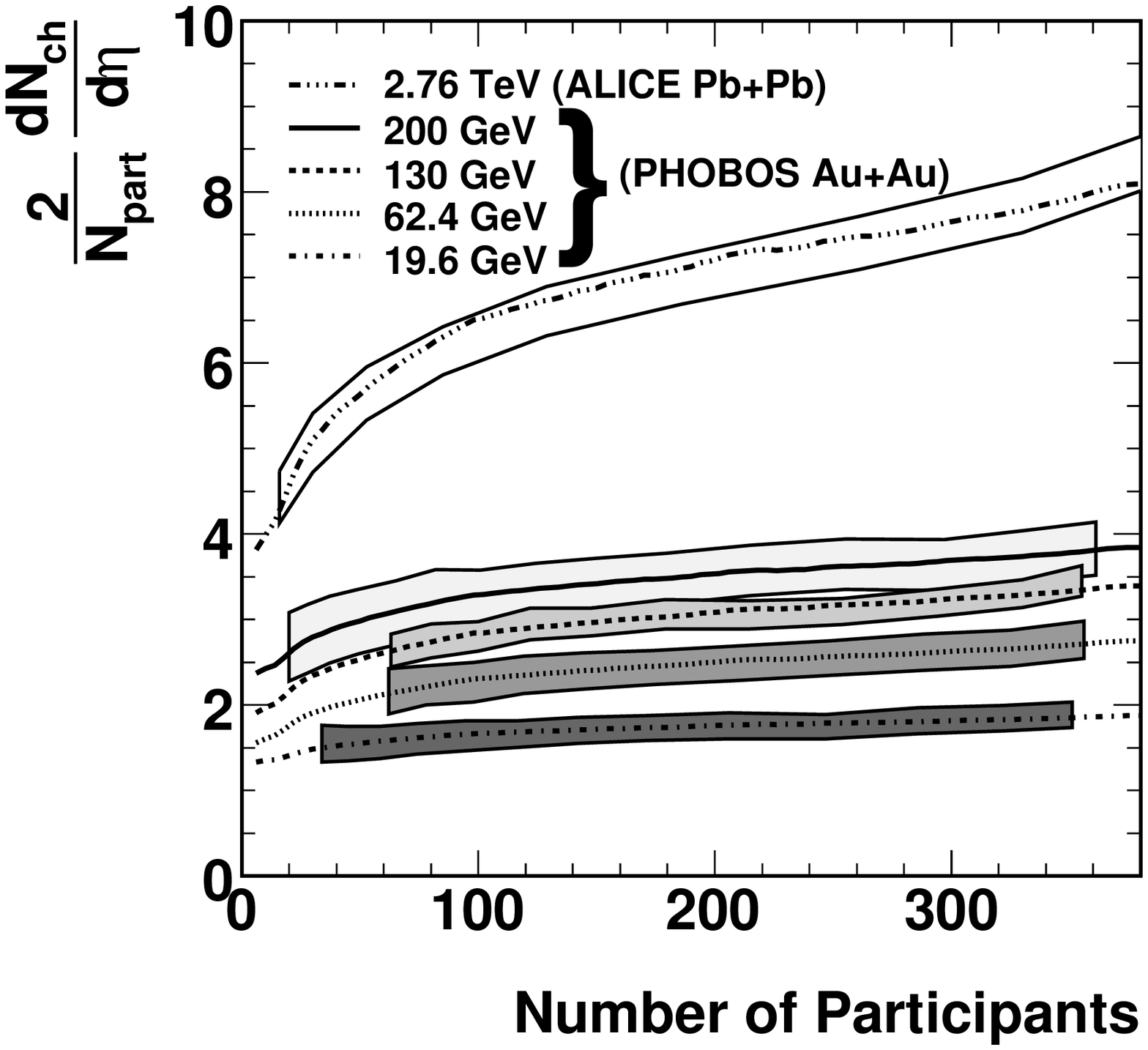}
\includegraphics[angle=0,width=0.375\textwidth]{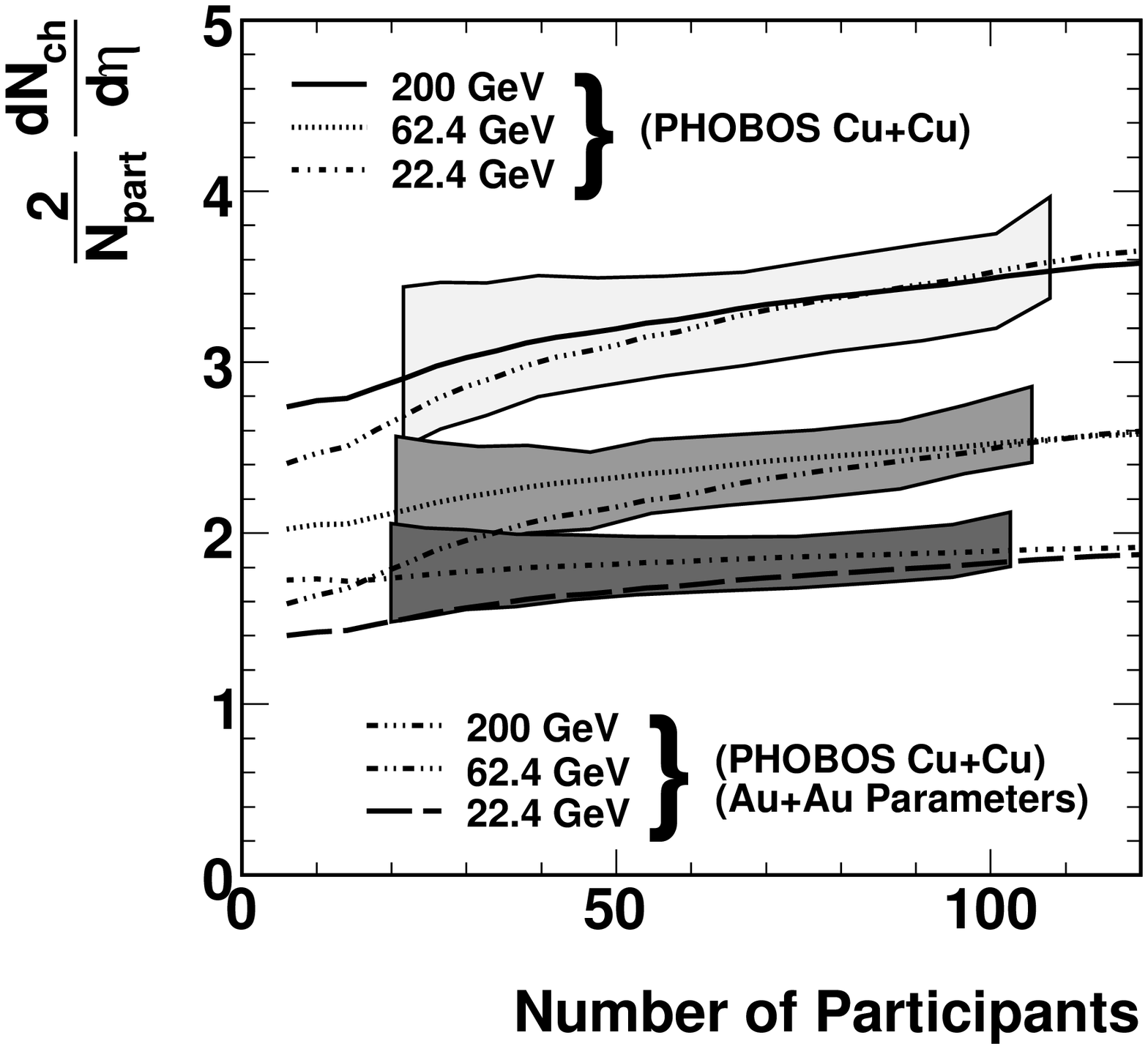}
\vspace{-20pt}
\caption{\label{fig:MultiplicityIImid}
PHOBOS~\cite{cite:PHOBOS_200mid,cite:PHOBOS_130mid,cite:PHOBOS_62.4mid,cite:PHOBOS_19.6mid,cite:PHOBOS_CuCu_MidRap} and ALICE~\cite{cite:ALICE_2.76mid} multiplicity data at mid-rapidity compared to that
derived from the fit, using the least-$\chi^{2}$ method,
for Au+Au (Pb+Pb) data in the left panel and Cu+Cu data in the right
panel.  Fits are shown for RHIC energies \snn\,=\,200\,GeV (lightest
grey band), 130\,GeV (light grey, Au+Au only)), 62.4\,GeV (dark grey), and
19.6\,GeV (22.4\,GeV for Cu+Cu) (darkest grey). The ALICE data (2.76\,TeV)
are shown as a no-fill outline.  For comparison, in the right panel, the
expected Cu+Cu data are shown using the \ymono~and \ymulti~yields found
from the fits to the Au+Au collision data.}
\end{figure*}

\begin{figure*}[ht]
\centering
\includegraphics[width=0.75\textwidth]{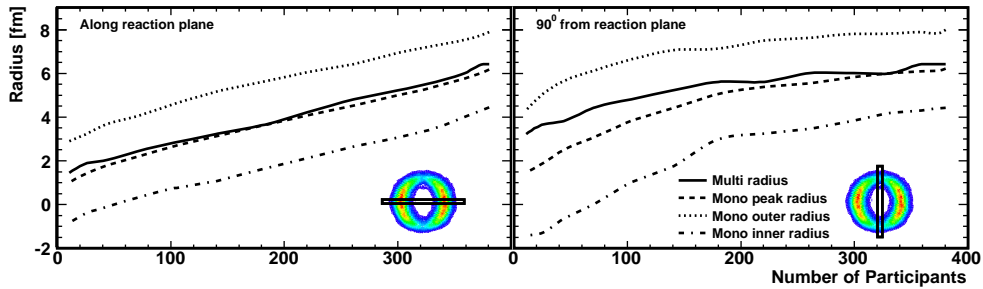}
\vspace{-18pt}
\caption{\label{fig:MeanRadii}
(color online) Mean mono and multi radii -- estimated from the Glauber
model.  The left (right) panel shows the radii along the ($90^{\circ}$ to)
reaction plane (the line along the nuclei centers -- $x$-
($y$-) axis in Fig.~\ref{fig:MonoMultiPositions}).  The solid
line shows the multi radius (defined as 10\% of the peak).
The different dashed lines represent the mono peak position
(dash) and the outer/inner radius, defined as 10\% of the peak
yield.  Negative radii (in the inner radius) represents the
point at which the mono are no longer distinctly separated by the
multi -- i.e. mono collisions occur at every point throughout the
collision area.}
\end{figure*}

\begin{table*}[th]
\caption{\label{tbl:MidRapVsEnergy} Mono and multi yields (and relative to
data \pp~inelastic
yields~\cite{cite:ppRef200,cite:PHOBOS_130mid,cite:ppRef62.4_22.4,cite:PHOBOS_19.6mid})
as derived from the least-$\chi^{2}$ fit to the
mid-rapidity data from PHOBOS (Au+Au and Cu+Cu) and ALICE (Pb+Pb).}
\begin{tabular}{| c || c | c || c | c || c | c || c | c |}
\hline
       & \multicolumn{4}{c ||}{Au+Au} & \multicolumn{4}{ c |}{Cu+Cu} \\
Energy & \multicolumn{2}{c ||}{$dN_{ch}^{mono}/d\eta$} & \multicolumn{2}{ c ||}{$dN_{ch}^{multi}/d\eta$} & \multicolumn{2}{c ||}{$dN_{ch}^{mono}/d\eta$} & \multicolumn{2}{ c |}{$dN_{ch}^{multi}/d\eta$} \\
(GeV)  & yield & yield/\pp & yield & yield/\pp & yield & yield/\pp & yield & yield/\pp \\\hline
 2760  & 1.35  &  --      & 8.55  &  --      &  --   &  --      &  --   &  --      \\
  200  & 1.49  & 0.65     & 4.01  & 1.75     & 2.07  & 0.91     & 3.83  & 1.67     \\
  130  & 1.04  & 0.46     & 3.56  & 1.74     &  --   &  --      &  --   &  --      \\
 62.4  & 0.86  & 0.45     & 2.88  & 1.50     & 1.58  & 0.82     & 2.75  & 1.43     \\
 19.6  & 0.99  & 0.78     & 1.94  & 1.52     & 1.53  & 1.20     & 1.98  & 1.56     \\\hline
\end{tabular}
\end{table*}

There is another way to extract the \ymono~and \ymulti~yields from the data.
Figure~\ref{fig:MultiplicityIImid} (left panel) shows the mid-rapidity Au+Au data from
PHOBOS~\cite{cite:PHOBOS_200mid,cite:PHOBOS_130mid,cite:PHOBOS_62.4mid,cite:PHOBOS_19.6mid} and
Pb+Pb data from ALICE~\cite{cite:ALICE_2.76mid} data (bands) together with a least-$\chi^{2}$ fit (lines)
using the \nmono~and
\nmulti~distributions from Fig.~\ref{fig:MonoMultiVsNpart}.  The right panel
of Fig.~\ref{fig:MultiplicityIImid} shows the Cu+Cu data from PHOBOS~\cite{cite:PHOBOS_CuCu_MidRap}
along with fits using that data and the extracted parameters from the fit to the Au+Au
collision data at the same energy.  (Note that the 22.4\,GeV Au+Au reference is 
scaled by 1.04 to account for the small difference in collision energy.)
This least-$\chi^{2}$ fit is a different approach than the pair fit method
used above.  This enables the full centrality dependence of the mid-rapidity
results to be simultaneously utilized and reduces the dependence of the 
results from the single pair choice.  The
resultant least-$\chi^{2}$ fit at each energy well represents the data, even over
two orders of magnitude of collision energy.  The derived mid-rapidity
mono and multi multiplicities
are summarized in Table~\ref{tbl:MidRapVsEnergy}.
For comparison,
these yields are shown as a star symbol to compared to the pair
fit method in Fig.~\ref{fig:MultiplicityII}.


In an attempt to reconcile the difference between the minimum bias
\pp~data and the derived underlying mono-distribution, we compare
the detailed spacial positions of the mono and multi interactions,
similar to those shown in Fig.~\ref{fig:MonoMultiPositions}.
Figure~\ref{fig:MeanRadii} shows the mean radius of the multi
(solid line) and mono (dot and dot-dashed lines) defined at
10\% of the peak.  The center of the mono (peak) is shown as the
dashed line, which coincides with the multi radius.  The
left (right) figures show the radii along ($90^{\circ}$ to) the
reaction plane.  Clearly, half of the mono distribution resides
within the multi distribution.  One could well imagine that if the
multi represents an opaque medium, then much of the mono inside
the multi region could be absorbed (or at least modified/suppressed).
Similarly, those outside the multi region may be suppressed should
the particles be headed toward the multi region, an eclipse-type
effect.

\begin{figure*}[t]
\centering
\begin{minipage}{0.24\textwidth}
\includegraphics[angle=0,width=1\textwidth]{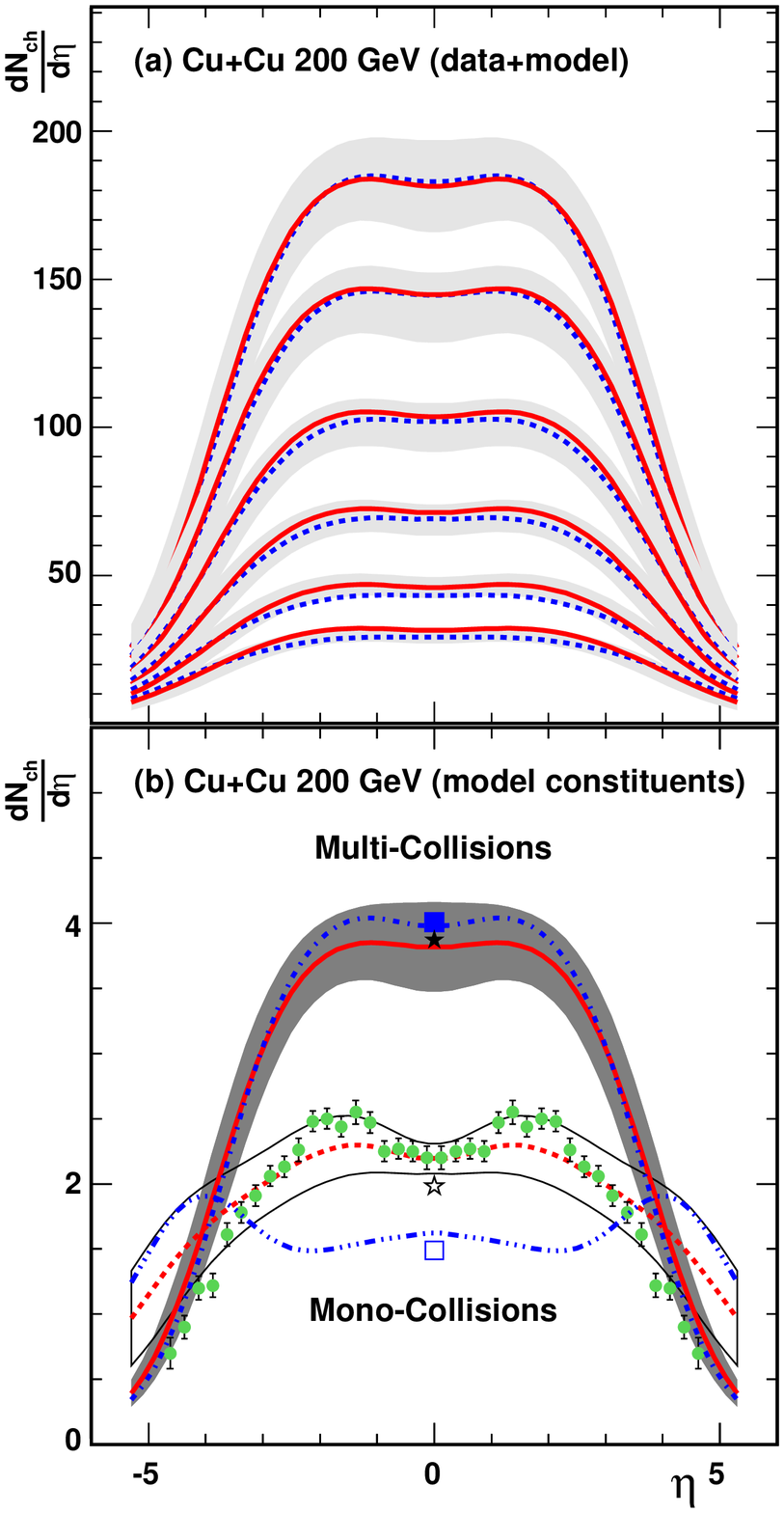}
\end{minipage}
\begin{minipage}{0.24\textwidth}
\includegraphics[angle=0,width=1\textwidth]{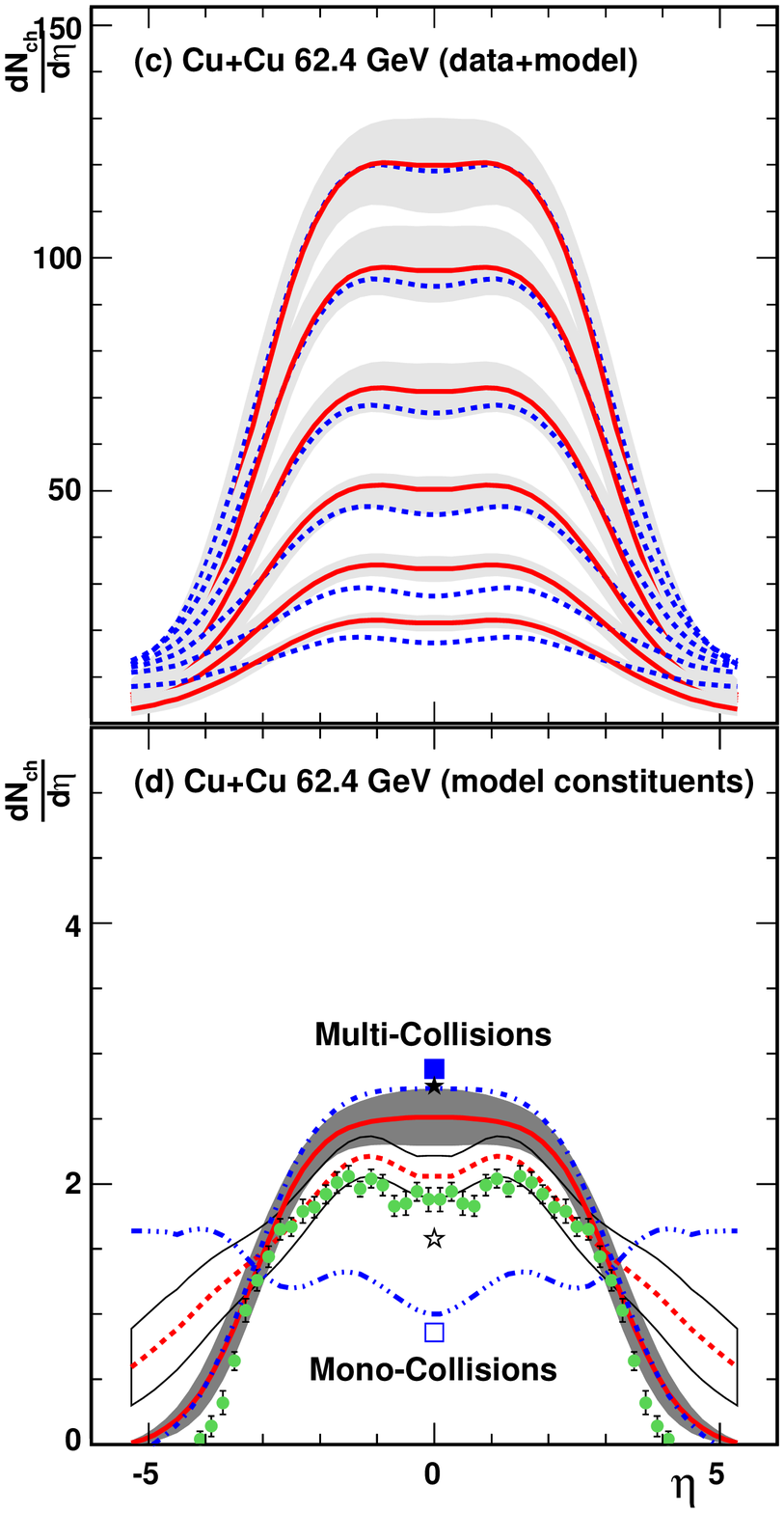}
\end{minipage}
\begin{minipage}{0.24\textwidth}
\includegraphics[angle=0,width=1\textwidth]{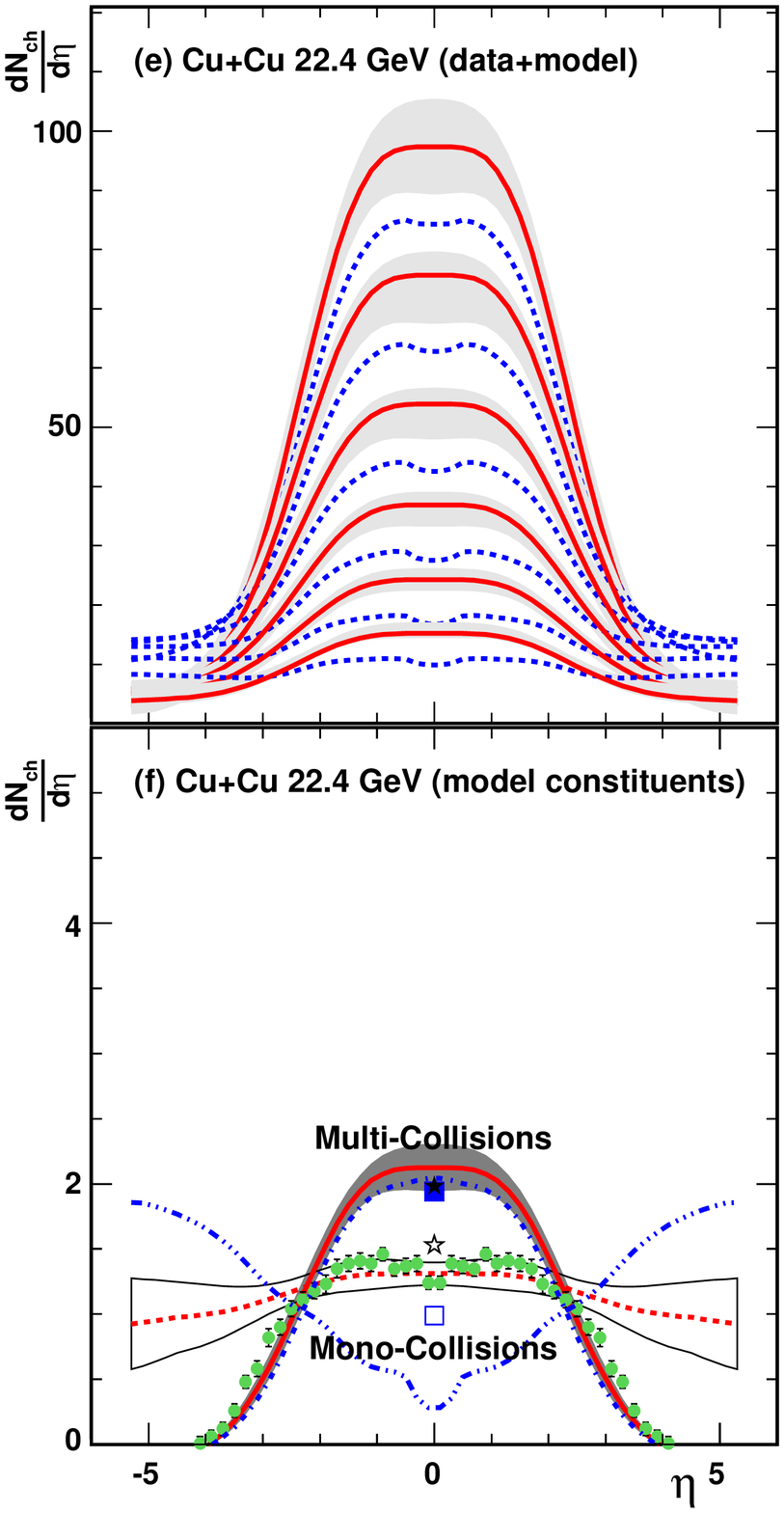}
\end{minipage}
\caption{\label{fig:MultiplicityIICuCu}
(color online) PHOBOS multiplicity data from Cu+Cu collisions~\cite{cite:PHOBOS_CuCudNdeta} compared
to the multiplicity distributions derived from the model, using the pair
fit method (upper panels, red) and from using the underlying \ymono~and
\ymulti~distributions from Au+Au collisions (blue, see
Fig.~\ref{fig:MultiplicityII}). Panels (a,b), (c,d), and (d,e) show
data for Cu+Cu collisions at 200, 62.4, and 22.4\,GeV respectively.
The lower panels show the underlying mono (dashed line/open band) and multi
(solid line/filled band) extracted from Cu+Cu data.  The minimum bias
\pp~data~\cite{cite:ppRef200,cite:ppRef62.4_22.4} (green
circles) and
the underlying mono and multi distributions from the Au+Au analysis (blue dashed lines) are
shown for reference.  The star (square) symbols represent the least-$\chi^{2}$
fit to the mid-rapidity Cu+Cu (Au+Au) data, see Fig.~\ref{fig:MultiplicityIImid} right (left) panel.}
\end{figure*}

One can test the model by using the Cu+Cu
data~\cite{cite:PHOBOS_CuCudNdeta} to extract new \ymono~and \ymulti~yields
from the data and compare those to the Au+Au underlying
mono and multi distributions.  Figure~\ref{fig:MultiplicityIICuCu}
shows a similar analysis as described above for the pair fit
method, this time with Cu+Cu collisions at \snn\,=\,200, 62.4, and
22.4\,GeV.  For completeness, the underlying \ymono~and \ymulti~distributions
from Au+Au collisions (blue dashed lines) are also used to
form the multiplicity data.
In comparing the underlying mono and multi distributions from Cu+Cu
data to those from Au+Au data, we find that the multi distributions 
are very similar and that the mono distribution in Cu+Cu more closely
resembles that from minimum bias \pp~interactions.  Similarly in the
right panel of Fig.~\ref{fig:MultiplicityIImid}, the same
least-$\chi^{2}$ method used for Au+Au collisions is applied to the 
mid-rapidity Cu+Cu data from PHOBOS~\cite{cite:PHOBOS_CuCu_MidRap} where 
the same trends are observed in terms of the underlying distributions.

\section{Unidentified Charged Hadron Spectra}

Following the success of the analysis of the charged particle
multiplicity, we apply this approach on more detailed data to check whether
a reasonable agreement could be found.  First, the unidentified charged hadron
spectra versus transverse momentum (\pT) is used.  This covers a
different kinematic region than the low-\pT~dominated multiplicity.
The suppression of high-\pT~particles in central events is a
fascinating phenomena which has been modeled by many different
theories, which are generally decoupled from the low-\pT~region.
Here, the full \pT~range of the data is used with the primary goal
of a qualitative
description of the data.  As an initial test, the \ymono~and \ymulti~are
again taken from \pythia, at a collision energy of 200\,GeV, with
mono derived from a truly minimum bias sample and
multi from several hypothesized distributions as described above.
Again, \nmono~and \nmulti~are from Fig.~\ref{fig:MonoMultiVsNpart}.

\begin{figure*}[h]
\centering
\includegraphics[angle=0,width=0.70\textwidth]{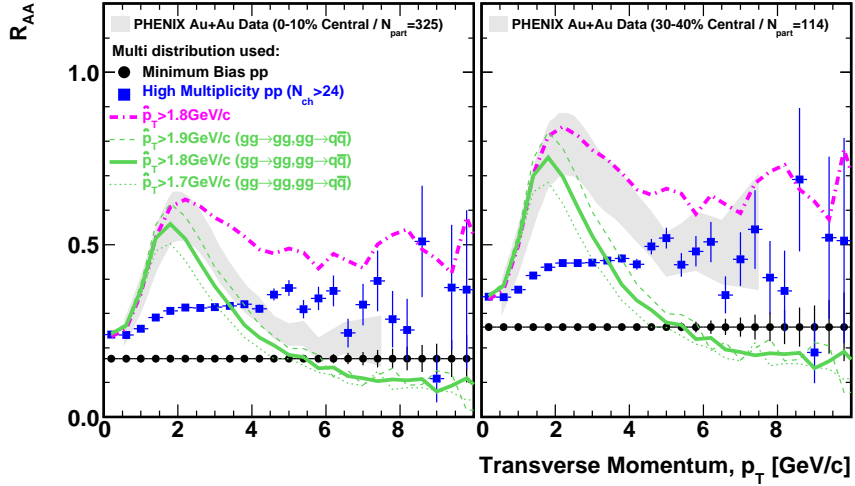}
\vspace{-24pt}
\caption{\label{fig:ChHadronSpectraI}
(color online) Comparison between measured nuclear modification factor,
\raa, versus \pT~(PHENIX data~\cite{cite:PHENIX_RAA_AuAu200}) (band)
and model predictions with various candidate \ymulti~distributions
(colored symbols) as noted in the legend.  0-10\% central (left) and
mid-central 40-50\% (right) data are shown.  In each model
representation, \ymono~is fixed to minimum bias
\pp~from \pythia.}
\end{figure*}

Figure~\ref{fig:ChHadronSpectraI} shows the nuclear modification factor,
\raa~-- see Eqn.~\ref{eqn:raa}, for Au+Au collisions at \snn\,=\,200\,GeV
for two centrality bins as measured by PHENIX~\cite{cite:PHENIX_RAA_AuAu200}.
It can be clearly seen that,
although the model distributions do not match the experimental
data precisely, the qualitative features of the data are reproduced.
Using the \pp~minimum bias (circles) or high multiplicity (squares)
simulations for the multi component of the hadron spectra do not
represent the data enhancement
in the intermediate \pT~region with either of these as the underlying multi
distribution.  The minimum bias result, in this case, represents a simple
visual scale of the number of collisions.

\begin{equation}
\label{eqn:raa}
\ensuremath{ \raa = \frac{1}{\ncoll}\frac{\frac{d^{2}N^{\rm Au+Au}}{dyd\pT}}{\frac{d^{2}N^{\pp}}{dyd\pT}}}
\end{equation}

\begin{equation}
\label{eqn:raanpart}
\ensuremath{ \raanpart = \frac{1}{\npart}\frac{\frac{d^{2}N^{\rm Au+Au}}{dyd\pT}}{\frac{d^{2}N^{\pp}}{dyd\pT}}}
\end{equation}

Applying a minimum $\hat{p}_{T}$ cut of 1.8\,GeV/$c$ on the minimum
bias sample (dot-dashed line) results in a clear peak close to the
minimum cut (as expected) but over predicts the higher-\pT~data.
Removing all quark interactions, leaving a gluon dominated system
(solid, dotted, and dashed lines) both
reproduces the intermediate-\pT~peak and the lower yield at
higher-\pT.  To emphasize again, the objective of this analysis is
not to {\it find} the underlying distribution, but to show that a
simple underlying distribution could exist which can describe the
features observed in heavy ion data.

\begin{figure*}[h]
\centering
\includegraphics[angle=0,width=0.90\textwidth]{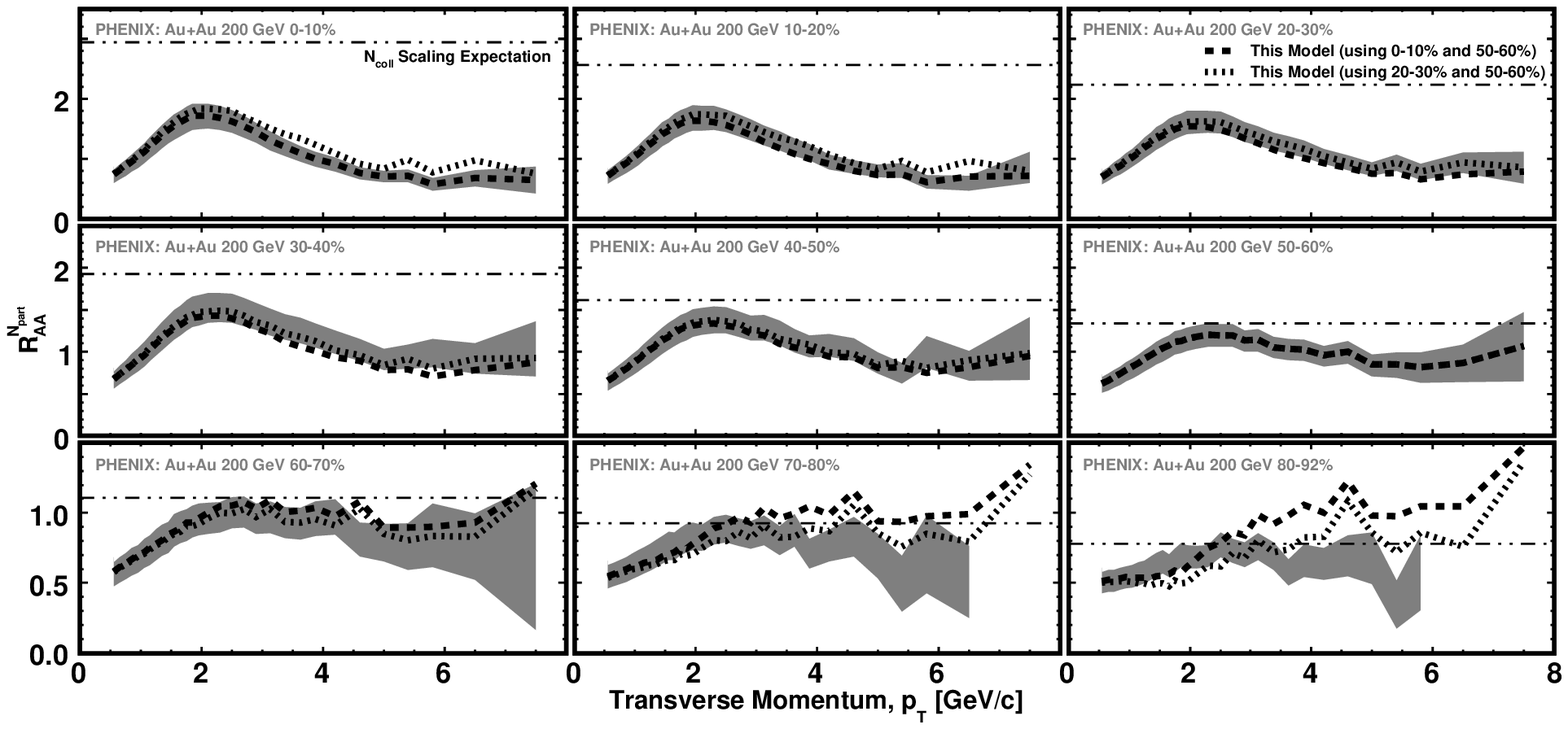}
\caption{\label{fig:ChHadronSpectraII}
Nuclear modification factor (scaled by \npart~not \ncoll) for
unidentified charged hadrons from PHENIX~\cite{cite:PHENIX_RAA_AuAu200}
(filled band) at \snn\,=\,200\,GeV.  The lines are the
\raanpart~reconstituted from the underlying mono- and multi-distributions
using the pair fit method; dashed (dotted) lines use centrality bins
0-10\% (20-30\%) and 60-70\%.  In each figure, the dot-dashed line
represents the expected \ncoll~scaling, assuming particle production
scales as \pp~$\times$~\ncoll.}
\end{figure*}

Using the same technique to fit the data as described above for the
multiplicity `pair fit', charged hadron spectra can be modeled
using PHENIX data~\cite{cite:PHENIX_RAA_AuAu200}. The
fit is performed twice, once using the 0-10\% and 60-70\%
cross-section bins as the data pair to obtain the \ymono~and
\ymulti~distributions and a second pairing of 20-30\% and 60-70\% to test
the sensitivity in the choice of bin pairing, see
Fig.~\ref{fig:ChHadronSpectraII}.
The extracted underlying mono and multi distributions are then used
to model \raanpart, Eqn.\ref{eqn:raanpart}, in the remaining centrality bins.

\begin{figure*}[h]
\centering
\includegraphics[angle=0,width=0.375\textwidth]{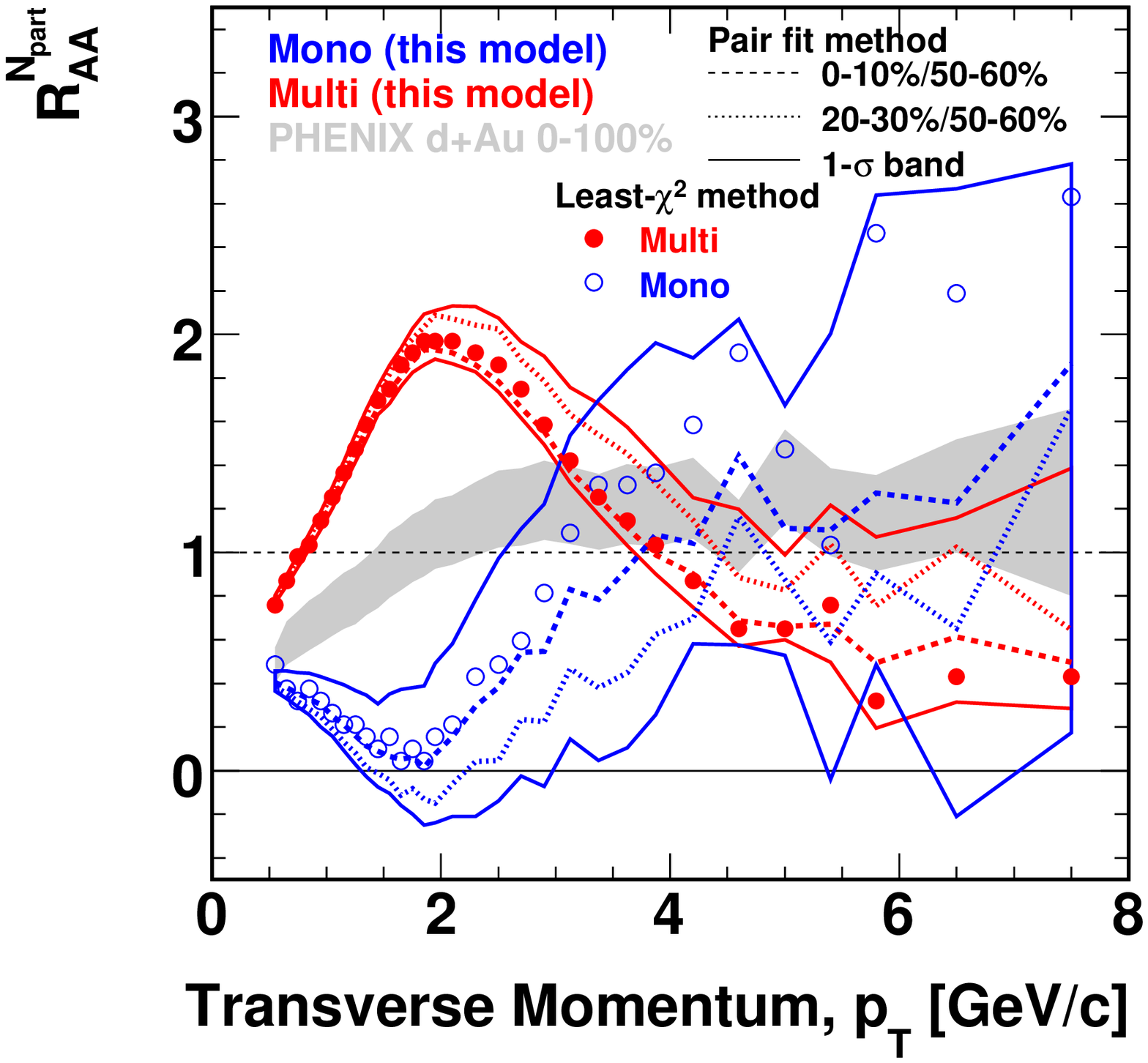}
\includegraphics[angle=0,width=0.375\textwidth]{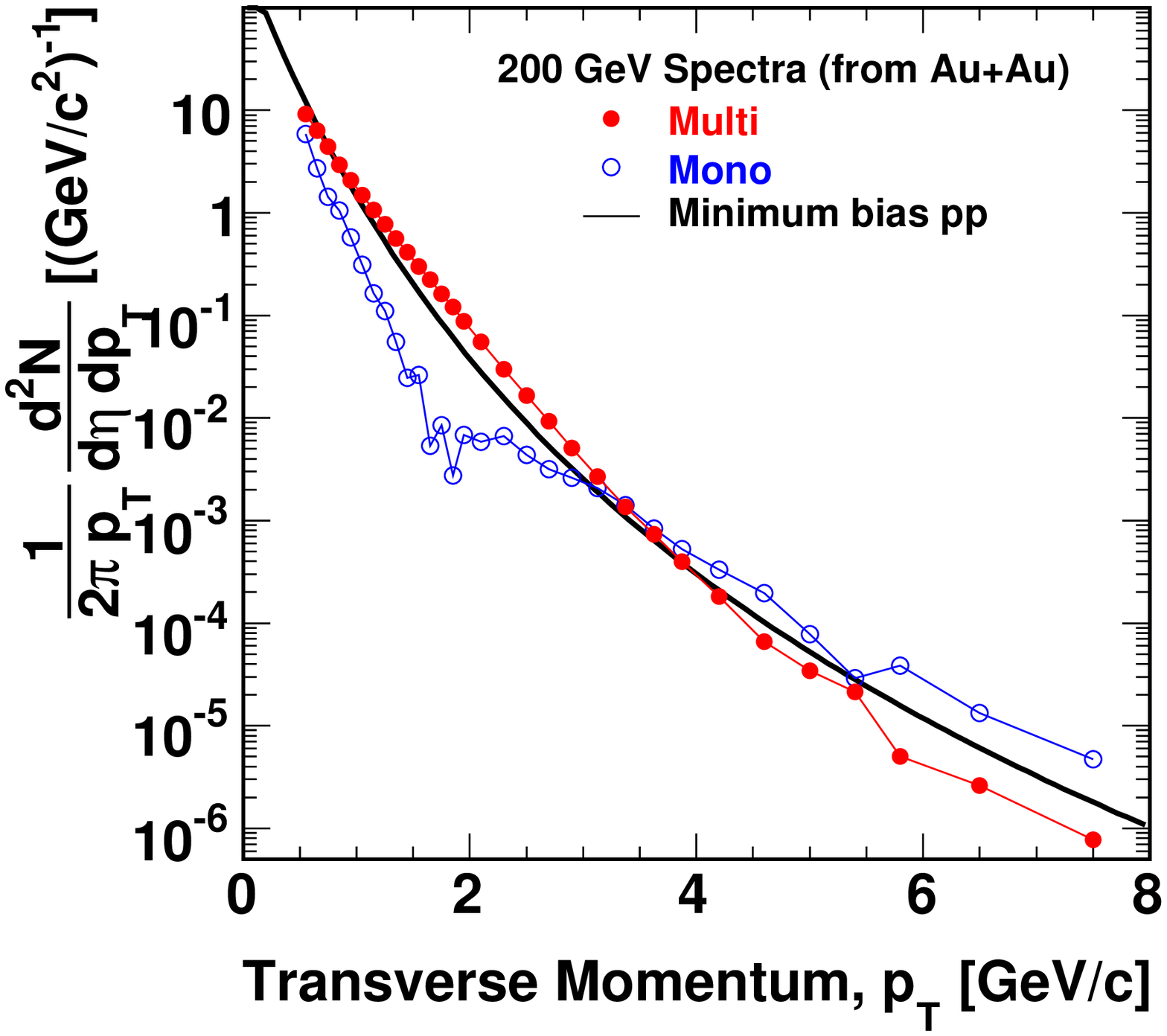}
\vspace{-20pt}
\caption{\label{fig:ChHadronSpectraIIresult}
(color online) Mono and multi nuclear modification factors (scaled by
\npart~not \ncoll) using the pair fit method on PHENIX,
PHOBOS~\cite{cite:PHOBOS_RAA_AuAu200}, and
STAR~\cite{cite:STAR_RAA_AuAu200} data.  The left panel shows
\snn\,=\,200\,GeV mono and multi distributions,
derived from a fit to the PHENIX data.  The dashed and dotted lines
show the fits from the centrality bins described in
Fig.~\ref{fig:ChHadronSpectraII}; the outline shows the 1-$\sigma$
band, derived from all possible centrality bin pairs. The grey band
shows equivalent PHENIX $d$+Au data~\cite{cite:PHENIX_dAu_Spectra}
from minimum bias collisions for reference.  The closed (open) circles
represent the results from the least-$\chi^{2}$ method applied to the
spectra (omitting the four lowest centrality points from the fit).
The right panel shows the mono and multi spectra using the least-$\chi^{2}$
fit method.  The closed red (open blue) circles represent the multi
(mono) collisions spectra per participant.  The solid line represents
the 200\,GeV pp spectra, taken from~\cite{cite:PHENIX_RAA_AuAu200}.}
\end{figure*}

\begin{figure}[h]
\centering
\includegraphics[angle=0,width=0.375\textwidth]{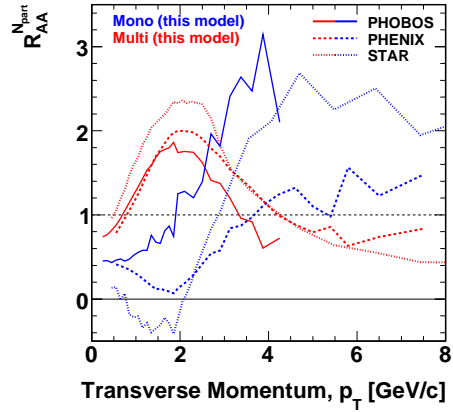}
\caption{\label{fig:ChHadronSpectraIIresultComp}
(color online) Comparison of the mean mono and multi distributions
derived from fits to PHOBOS (solid lines), PHENIX
(dashed), and STAR (dotted) for Au+Au collisions at 200\,GeV.}
\end{figure}

We find that all multi-dominated centrality bins are well reproduced
from the underlying mono and multi distributions using any two bins;
the results are not sensitive to the choice of centrality bins.  In
each figure, the double-dot dashed line represents the `\ncoll'
scaling.  For data presented as
\raa, the line would be exactly at unity and the data yield (at
high-\pT) would represent the apparent suppression.  Within the mono/multi framework, we find that suppression at 
high-\pT~is not necessary to describe the data.  The most
peripheral bins, perhaps, are not as well described.  An explanation
for this limited agreement could be that a ``collectivity'' which
produces the multi distribution could be dissipating.  This could
point to a possible need to modify the model to account for an
additional scaling variable for example mono, duo and multi$^{\ge 3}$;
this is outside the scope of the current paper.

\begin{figure*}[h]
\centering
\includegraphics[angle=0,width=0.90\textwidth]{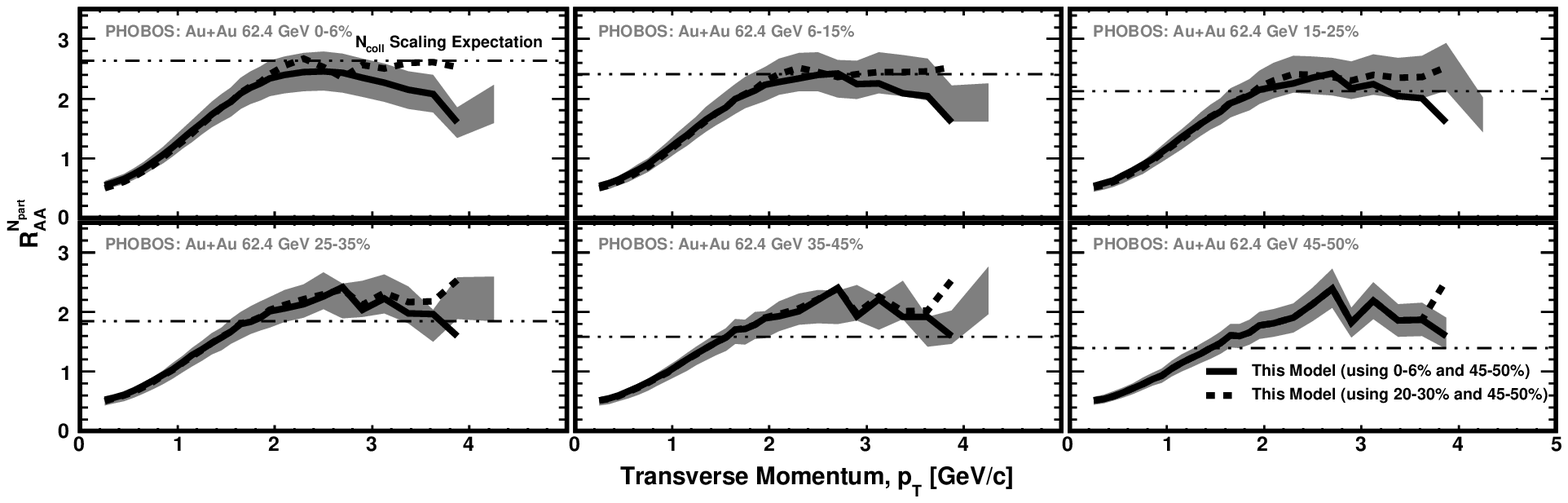}
\caption{\label{fig:ChHadronSpectraII62}
Nuclear modification factor (scaled by \npart~not \ncoll) for
unidentified charged hadrons from PHOBOS~\cite{cite:PHOBOS_RAA_AuAu62}
(filled band) at \snn\,=\,62.4\,GeV.  The lines are the
\raanpart~reconstituted from the underlying mono- and multi-distributions
using the pair fit method; solid (dashed) lines use centrality bins
0-6\% (15-25\%) and 45-50\%.  In each figure, the dot-dashed line
represents the expected \ncoll~scaling, assuming particle production
scales as \pp~$\times$~\ncoll.}
\end{figure*}

Figure~\ref{fig:ChHadronSpectraIIresult} (left panel) shows the
underlying mono and multi nuclear modification factors for \snn\,=\,200\,GeV
derived from the pair fit method; with the corresponding spectra in the 
right panel.  The dashed and dotted lines
represent the centrality bin pairs used in the analysis shown in
Fig.~\ref{fig:ChHadronSpectraII}.  The red outline shows a
1-$\sigma$ band around the mean underlying multi distribution
determined from an average over all possible centrality-pair
combinations, except the most peripheral.  This appears to be
well constrained, illustrating only a small variance in the
underlying multi distribution.  A comparative analysis using the
least-$\chi^{2}$ method (for the yields versus centrality for each
\pT~bin) was also performed.  (Note that the lowest four centrality
points were omitted from the fit.)  The results are shown as circles in
Fig.~\ref{fig:ChHadronSpectraIIresult} (left panel) and are in
good agreement with the distributions extracted from the pair fit
method.  The extracted underlying mono distribution, however, is less
constrained, possibly due to a changing surface suppression of
mono with centrality.  Nonetheless,
the extracted values of the mono distributions here are
in agreement with the ones obtained from fits to the Au+Au
charged particle multiplicity: at very low \pT, a suppression
is observed with respect to minimum bias \pp~collisions.
In the intermediate \pT~region the mono distribution falls to zero,
hinting that the absorption (or suppression) is maximal in this
kinematic region.  For some fits, it is found that the suppression
is `negative' this would represent a case when some additional
suppression is present from the edge of the multi-region; pointing,
perhaps, toward the need for a mono, duo, and multi$^{\ge 3}$ prescription.
Figure~\ref{fig:ChHadronSpectraIIresultComp} shows a comparison of
the resultant distributions by using PHENIX,
PHOBOS~\cite{cite:PHOBOS_RAA_AuAu200}, and
STAR~\cite{cite:STAR_RAA_AuAu200} data.  The multi
distribution in this analysis are found to be very similar.
To put the results into context, Fig.~\ref{fig:ChHadronSpectraIIresult}
shows the multi spectra (per participant) relative to the minimum bias
\pp~spectra, which is found to have the same systematic dependencies as
the gluon-dominated system found in Fig.~\ref{fig:ChHadronSpectraI}.
It has been seen that, through the separation into multi and mono
sub-components of the collision, a consistent picture can be observed.
At the center of this picture is a dense (perhaps gluon dominated) system
which is opaque to the particles produced at the periphery of the
collision.  A gluon-dominated picture is not excluded by other
measurements, for example, the anomalous increase of baryons, with
respect to mesons, could be expected from a system with a higher number
of gluon-jets than quark-jets.

\begin{figure*}[h]
\centering
\includegraphics[angle=0,width=0.375\textwidth]{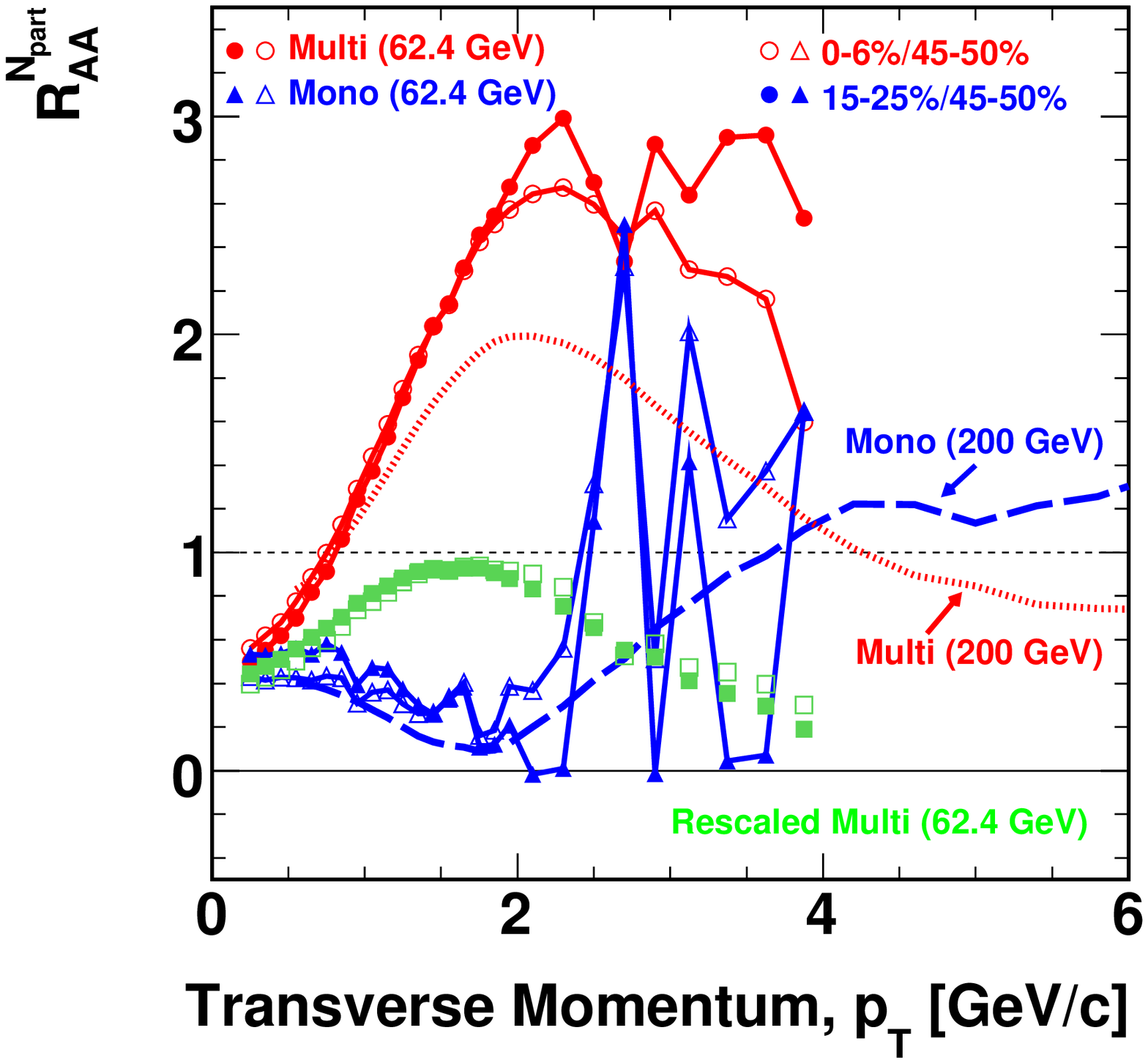}
\includegraphics[angle=0,width=0.375\textwidth]{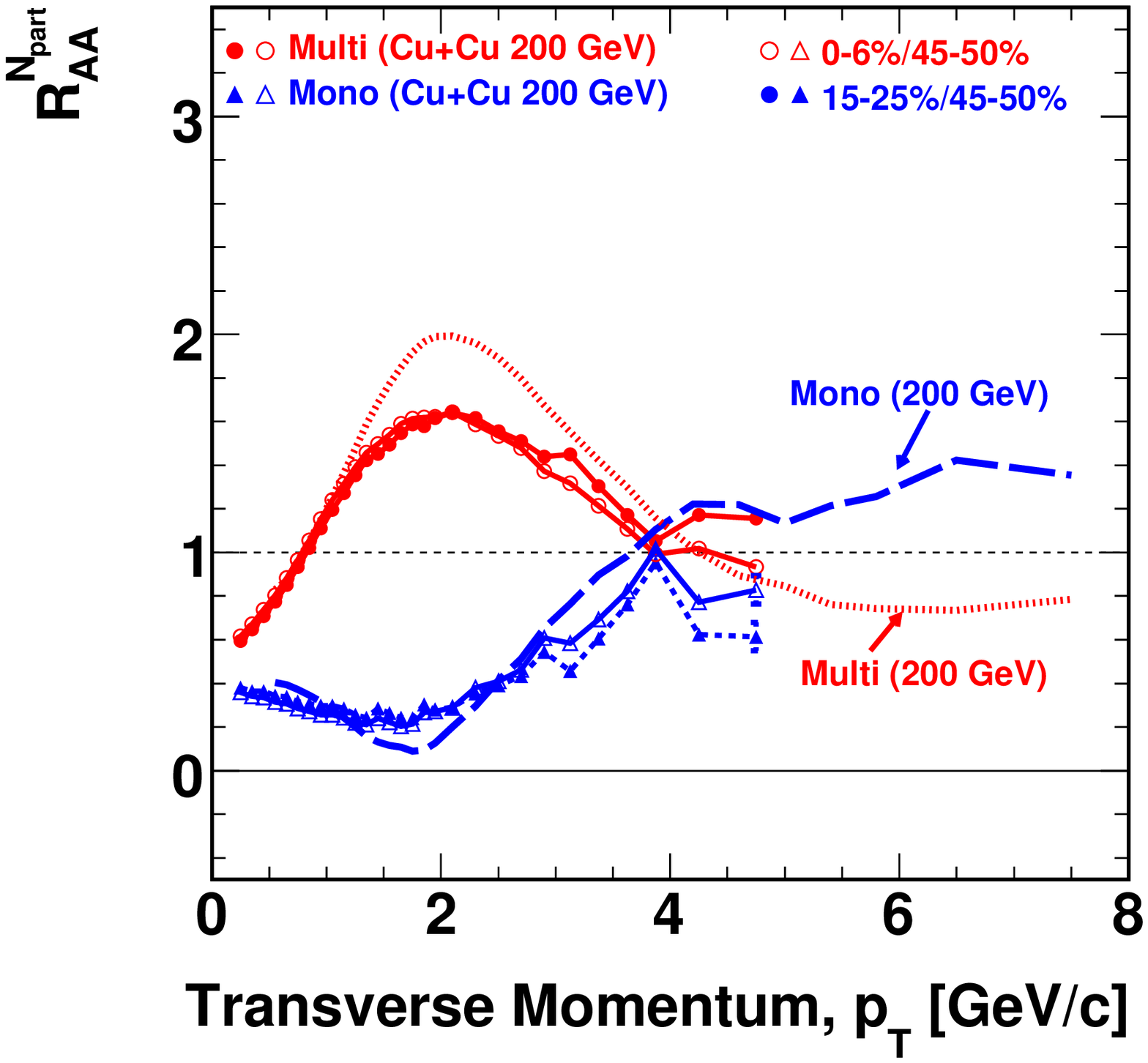}
\caption{\label{fig:ChHadronSpectraIIresult2}
(color online) 
The left panel shows the underlying mono (triangles) and multi (circles)
\raanpart~distributions for PHOBOS Au+Au data at
\snn\,=\,62.4\,GeV~\cite{cite:PHOBOS_RAA_AuAu62}.  The solid lines are
meant to guide the eye.  The open (closed) symbols represent a pair fit
using centrality bins 0-6\% and 45-50\% (0-6\% and 45-50\%).
The dashed lines depict the underlying mono and multi distributions
from Au+Au collisions at 200\,GeV (PHENIX).  In order to better compare
the multi distributions, the green squares represent the 62.4\,GeV data
rescaled to the same \pp~reference (i.e. assumes a 200\,GeV \pp~reference,
not 62.4\,GeV).
The right panel has the same notation as the left, except the data is for
PHOBOS Cu+Cu collision data at 200\,GeV~\cite{cite:PHOBOS_RAA_CuCu}.}
\end{figure*}

It is interesting to note that the disappearance of the away-side
jet in two-particle correlations was observed in a similar
\pT~region (for the associated particles) as the mono
suppression~\cite{cite:STAR_DisappB2BJet}.
In this kinematic region, 1$<$\pT$<$3\,GeV/$c$, we find that
the underlying mono distribution is very small which perhaps hints
that jets are dominantly from the mono (\pp-like) interactions which
occur on the surface.  With this assumption, the away-side is then
suppressed (or fully absorbed) within the (multi) medium.  At higher
\pT, the mono and multi distributions essentially become the same and
are similar in magnitude to the measured Cronin-enhanced $d$+Au
data~\cite{cite:PHENIX_dAu_Spectra}.  It should be noted that in order
to form these spectral distributions, the \ncoll~scale is explicitly
removed.  \raanpart~distributions are formed entirely with our
participant-like parameters, thus any large-scale suppression in the
data (using \raa) is actually brought about by scaling
the distributions by \ncoll.  In other words, the suppression at
high-\pT~could be entirely an artifact of using the wrong scaling
variable (namely \ncoll).  How could \ncoll~be the wrong variable?
Consider a simple picture.  In a standard Glauber model calculation,
we assume that each nucleon-nucleon interaction is independent
and that the cross-section for each interaction is the same (i.e. 
$\sigma_{NN}$ does not change -- even after 10 collisions).
Once a ``collision'' has occurred one
could surmise that the cross-section is not the same, something has
been lost so the cross-section could be diminished.  Further, if one
assumes that the multi component
melts into a single (not discrete) system, then it may be impossible,
in fact improbable, to manifest multiple collisions, or at least
reliably count them.  Note that, experimentally, one cannot count
\ncoll, yet one can directly count \npart, or at least derive it
from the data with some certainty.

It has already been observed that by dividing the \snn\,=\,200\,GeV
charged hadron spectra by that from 62.4\,GeV, the ratio (for a given
\pT) is invariant across all centrality bins
measured~\cite{cite:PHOBOS_CuCu_MidRap}.  This may already indicate
a simple geometry scaling which factorizes in energy and centrality.
We should therefore expect a similarly successful fit using the mono/multi
approach for 62.4\,GeV
data.  Fig.~\ref{fig:ChHadronSpectraII62} shows the same analysis
using PHOBOS charged hadron spectra at \snn\,=\,62.4\,GeV~\cite{cite:PHOBOS_RAA_AuAu62}, where two
centrality bin pairs can be used to predict all other bins.
Fig~\ref{fig:ChHadronSpectraIIresult2} shows the resultant mono
distribution, illustrating the same level of suppression as 200\,GeV,
whereas 62.4\,GeV has a higher multi \raanpart.  In the calculation
of \raanpart, the \pp~spectrum is used in the denominator to observe
any differences between A+A and \pp~collisions.  To more fairly compare
the two spectra, the green squares represent the same 62.4\,GeV data,
but rescaled to effectively use the \pp~200\,GeV reference.  In this 
way, we see that the underlying multi spectrum grows with energy.
A similar rescaling of the underlying mono distribution is not performed
as for that case, the relative change to \pp~is important, not the
overall spectrum.  It is interesting to note that \raanpart~for
mono-collisions is the same for 200\,GeV and 62.4\,GeV.  This
indicates that the level of suppression does not depend on collision
energy (at least for these RHIC energies).

\begin{figure*}[!h]
\centering
\includegraphics[angle=0,width=0.90\textwidth]{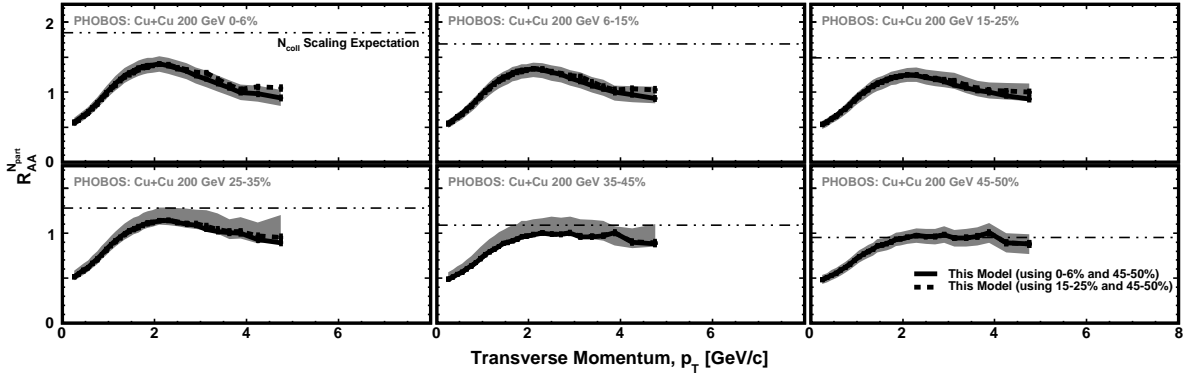}
\caption{\label{fig:ChHadronSpectraIICuCu}
Nuclear modification factor (scaled by \npart~not \ncoll) for
unidentified charged hadrons from PHOBOS (filled band) in Cu+Cu
collisions at \snn\,=\,200\,GeV~\cite{cite:PHOBOS_RAA_CuCu} and
from the model fit using the pair fit method (lines).  For the
solid lines, the centrality
bins 0-6\% and 45-50\% were used, dashed lines used centrality
bins 15-25\% and 45-50\%.}
\end{figure*}

\begin{figure*}[!h]
\centering
\includegraphics[angle=0,width=0.90\textwidth]{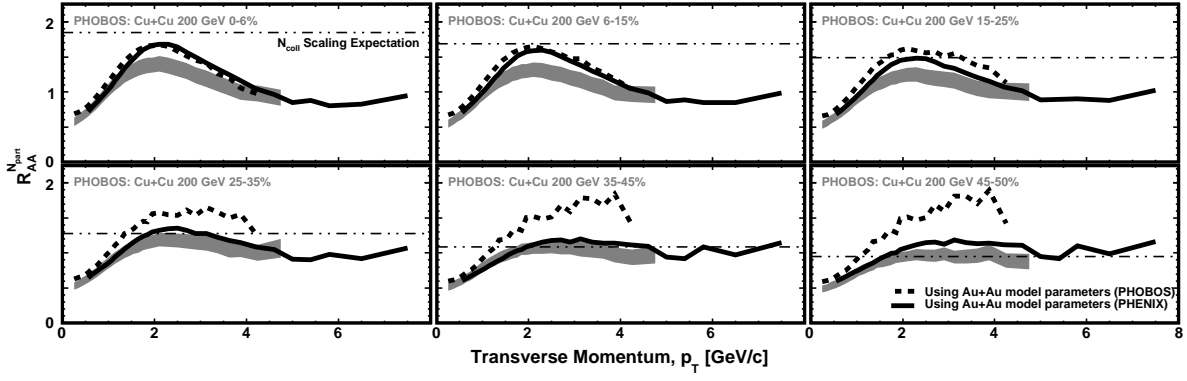}
\caption{\label{fig:ChHadronSpectraIICuCu_AuAuParams}
Nuclear modification factor (scaled by \npart~not \ncoll) for
unidentified charged hadrons from PHOBOS (filled band) in Cu+Cu
collisions at \snn\,=\,200\,GeV using the model fit results for
the mono and multi distributions from the Au+Au analysis.
The dashed (solid) lines use the results from fitting the PHOBOS
(PHENIX) data.}
\end{figure*}

In the multiplicity analysis, we tested the hypothesis that
common underlying distributions for the Au+Au and Cu+Cu data can be used
to represent the data.
It is possible to also perform the same comparisons with the unidentified
charged hadron spectra.  In Fig.~\ref{fig:ChHadronSpectraIICuCu}, a pair-fit
using PHOBOS Cu+Cu data at \snn\,=\,200\,GeV~\cite{cite:PHOBOS_RAA_CuCu}
is invoked to extract the underlying mono- and multi-distributions,
which are then used to reconstitute the \raanpart~in Cu+Cu for each centrality
bin.  In contrast, Fig.~\ref{fig:ChHadronSpectraIICuCu_AuAuParams}
uses the underlying mono- and multi-distributions derived from Au+Au
collisions to form the \raanpart~for each centrality bin.
In both cases, a reasonable fit is found.  For completeness,
the right panel in Fig.~\ref{fig:ChHadronSpectraIIresult2} shows a
comparison between the underlying mono and multi distributions
from Au+Au (symbols) and those derived from Cu+Cu (lines).  The
two sets of symbols for Cu+Cu represent the pair fit method using
PHOBOS centrality bins 0-6\% and 45-50\% (open symbols) and 0-6\%
and 45-50\% (closed symbols).

\section{Freeze-out Properties}
\begin{figure*}[!bh]
\begin{minipage}{18pc}
\includegraphics[angle=0,width=0.95\textwidth]{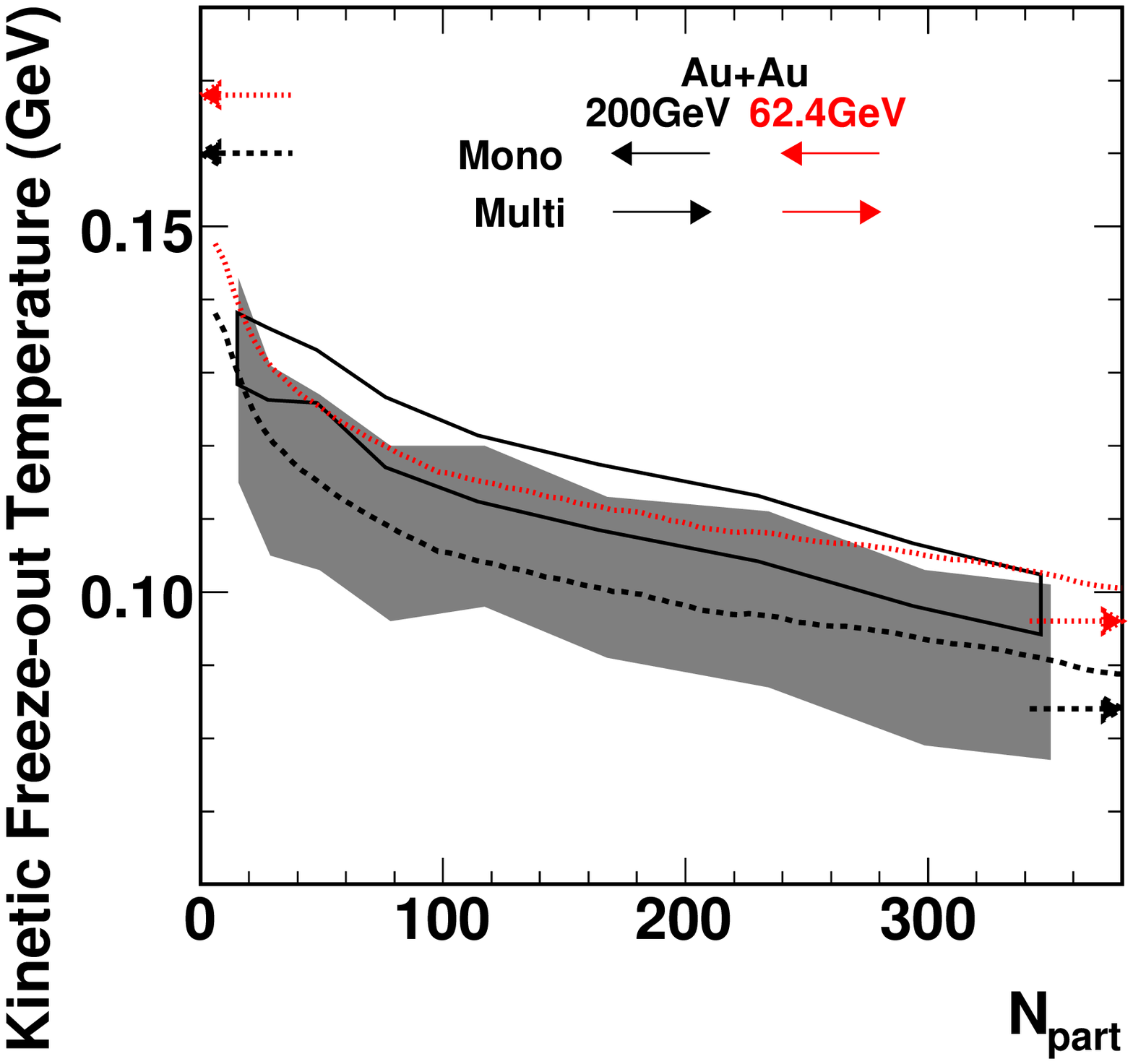}
\end{minipage}\hspace{2pc}
\begin{minipage}{18pc}
\includegraphics[angle=0,width=0.95\textwidth]{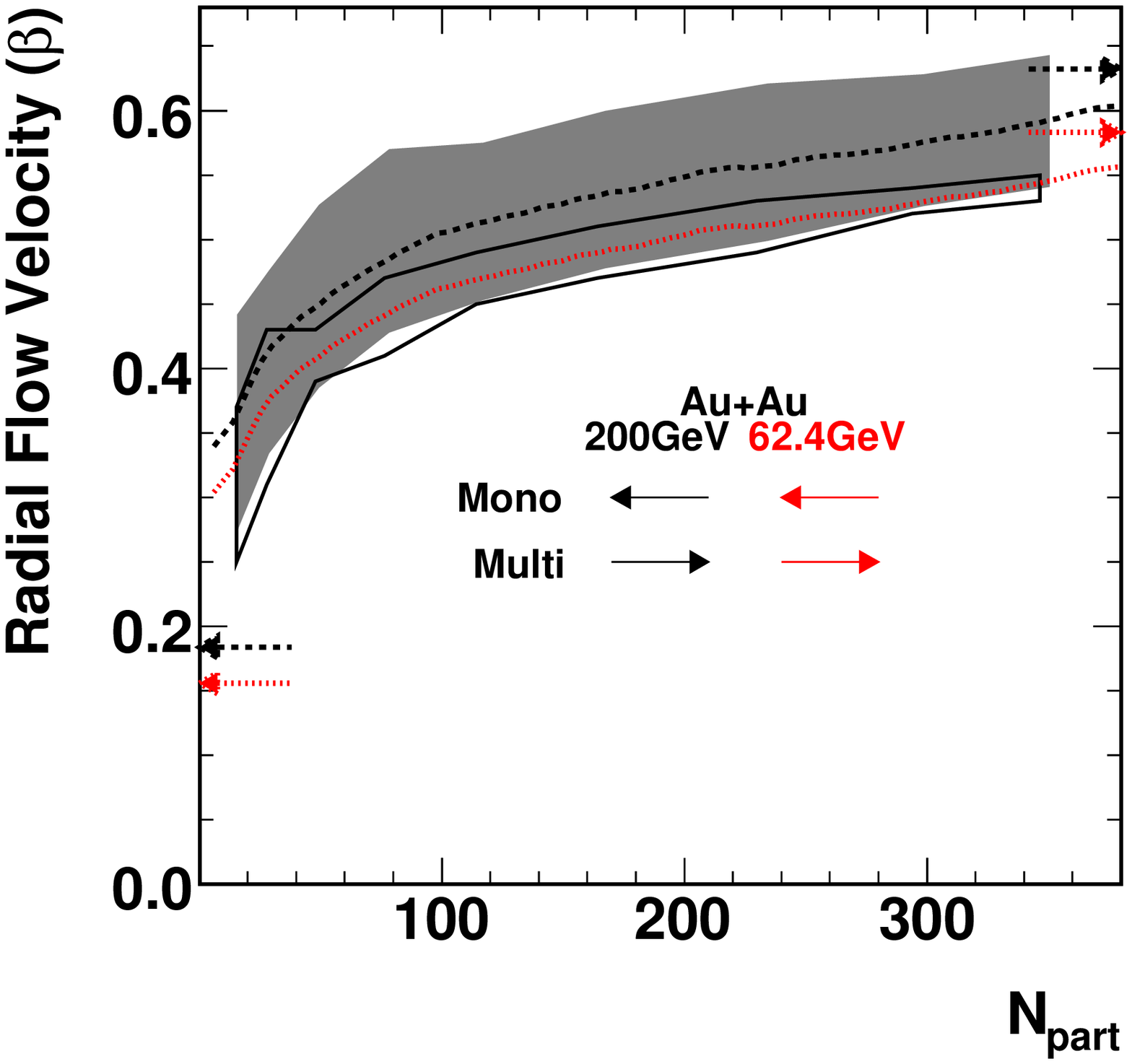}
\end{minipage}\hspace{2pc}
\caption{\label{fig:KinFO}
(color online) Least-$\chi^{2}$ fit to the kinetic freeze-out temperature
(left) and radial flow velocity (right) for Au+Au collisions at 200 (filled
band) and 62.4\,GeV (outline band).  Dashed (dotted) lines represent the
200\,GeV (62.4\,GeV) fit results.  The left (right) pointing arrows depict
the mono (multi) $T_{\rm kin}$ or $\beta$ in each panel.}
\end{figure*}

As an extension to the low-\pT~particle studies, it has proven
useful to fit the resultant spectra with models to derive freeze-out
properties of the system.  Commonly, the kinetic freeze-out
temperature ($T_{\rm kin}$) and radial flow velocity ($\beta$) of the
system are derived from a Blast-wave~\cite{cite:BlastWave} fit to
the low-\pT~spectra of identified pions, kaons, and protons.  For a
description see for example
Refs.~\cite{cite:STAR_CuCu_FreezeOut,cite:STAR_AuAu_FreezeOut}.
At kinetic freeze-out all elastic collisions cease and the spectral
shape is fixed.  The extracted parameters reflect the state of the
system at that time.
What has been observed, through the Blast-wave fits, is that
$T_{\rm kin}$ and $\beta$ vary quite strongly as a function of
centrality.  Here, we address the question whether those results
can be described within the mono/multi framework.  In a similar way
to the above analysis, we consider that mono and multi have their
own ``universal'' $T_{\rm kin}$ and $\beta$ values.  Using the
least-$\chi^{2}$ method described above, we fit the centrality
dependence of the freeze-out variables and extract those parameters
($Y$ variables in Eqn.\ref{eqn:MonoMulti}) for mono and multi.
The left and right panels of Fig.~\ref{fig:KinFO} shows the result of
fitting the freeze-out temperature ($T_{\rm kin}$) and radial flow velocity
($\beta$), respectively, at \snn\,=\,200 and 62.4\,GeV from STAR~\cite{cite:STAR_AuAu_FreezeOut}.  From the fits we find
that the centrality dependence is well described in this framework,
and the extracted $T_{\rm kin}$ and $\beta$ values for the underlying
mono and multi
components are given in Table~\ref{tbl:FreezeOut}.  Although there
is some collision energy dependence, the more striking feature is
that the mono results for $T_{\rm kin}$ and $\beta$ are significantly
different from the multi values.  A consistent picture emerges
that the multi component has considerably larger $\beta$
values and a freeze-out temperature of about one-half of that of the mono.

\begin{figure*}[!bh]
\begin{minipage}{18pc}
\includegraphics[angle=0,width=0.95\textwidth]{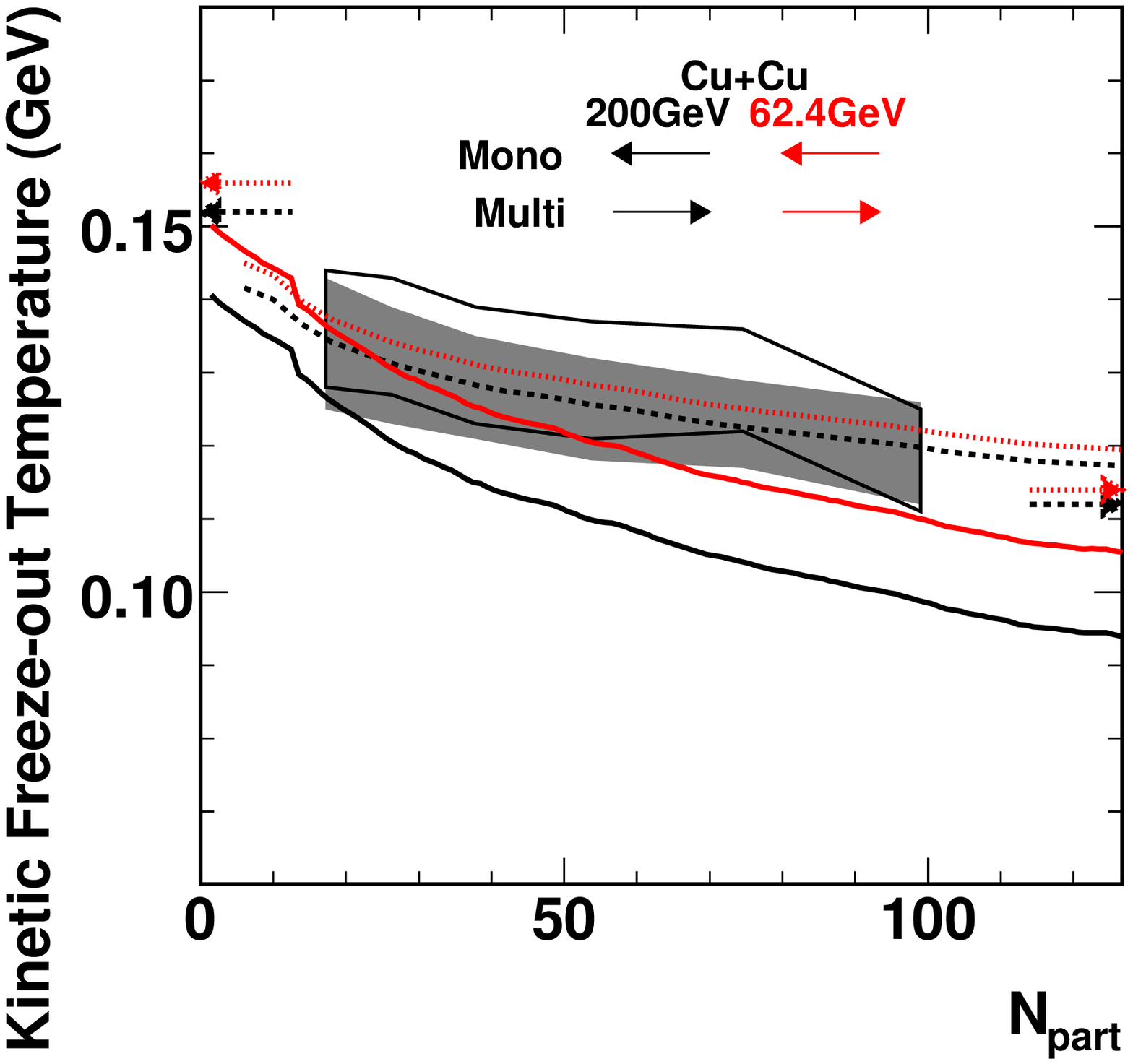}
\end{minipage}\hspace{2pc}
\begin{minipage}{18pc}
\includegraphics[angle=0,width=0.95\textwidth]{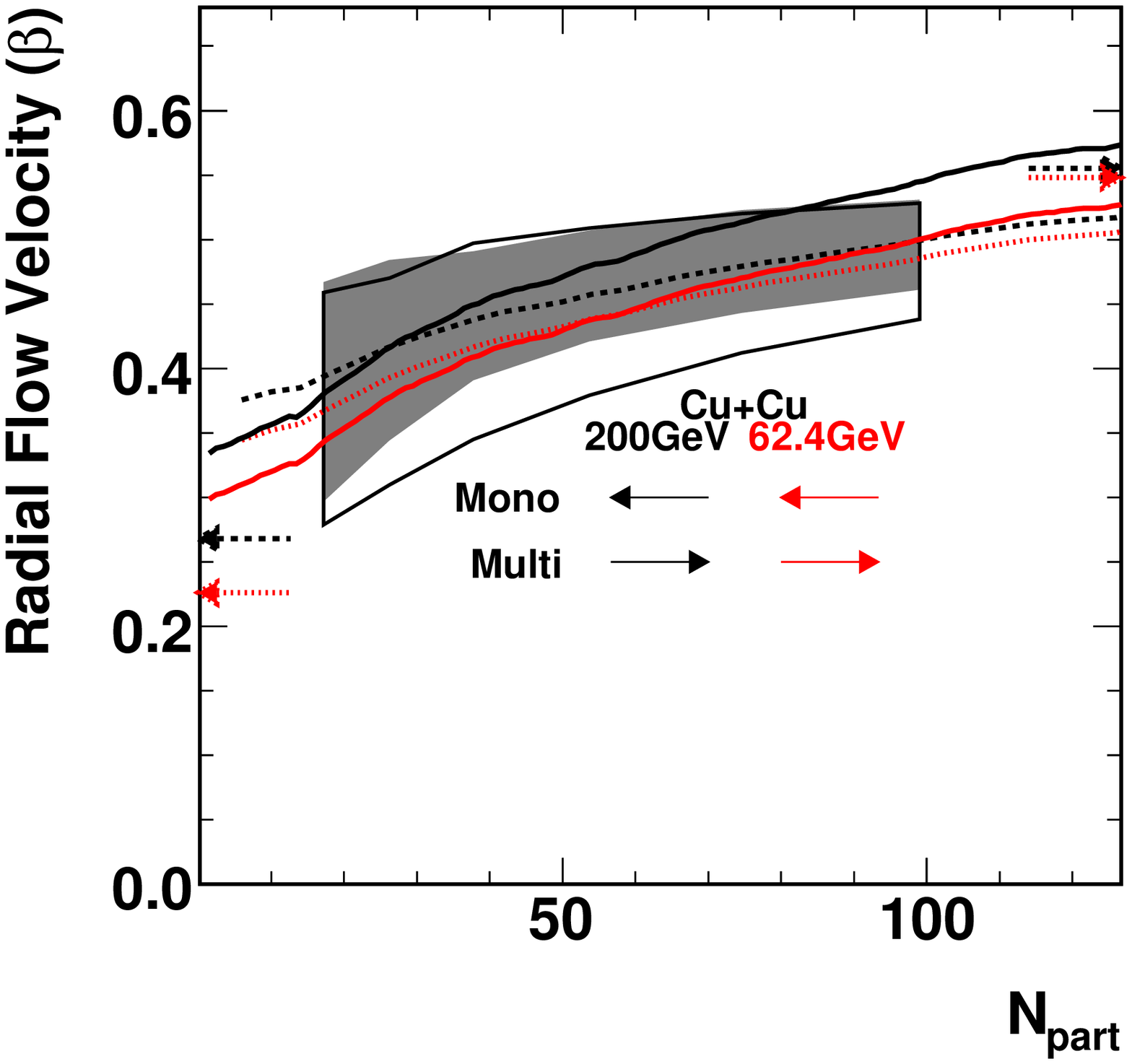}
\end{minipage}\hspace{2pc}
\caption{\label{fig:KinFOCuCu}
(color online) Least-$\chi^{2}$ fit to the kinetic freeze-out temperature
(left) and radial flow velocity (right) for Cu+Cu collisions at 200 (filled
band) and 62.4\,GeV (outline band).  Dashed (dotted) lines represent the
200\,GeV (62.4\,GeV) fit results.  The left (right) pointing arrows depict
the mono (multi) $T_{\rm kin}$/$\beta$.}
\end{figure*}

Figure~\ref{fig:KinFOCuCu} shows the results of a comparative analysis
using the freeze-out parameters from Cu+Cu collisions at
STAR~\cite{cite:STAR_CuCu_FreezeOut}.
Differences between Au+Au and Cu+Cu evident, especially
at the highest energy.  The heavier Au+Au system has a larger $\beta$
and smaller $T_{\rm kin}$ than those extracted from Cu+Cu collisions, 
see Fig.~\ref{fig:KinFOCuCu}.  Although there are differences, we
did not consider the systematic uncertainties in the data when creating
the underlying distributions for the two collision systems, so it becomes difficult to draw any strong
conclusions.  In comparison to the freeze-out
parameters extracted for \pp~collisions, for Au+Au collisions at
200\,GeV, the extracted $T_{\rm kin}$ for the underlying mono component
is about 25\% higher than that from the corresponding \pp~value
(0.127$\pm$0.013~\cite{cite:STAR_AuAu_FreezeOut}).  Similarly, the
extracted radial flow velocity is about 35\% lower than the
corresponding \pp~value (0.244$\pm$0.081~\cite{cite:STAR_AuAu_FreezeOut}).
We should note here, that we do not expect the mono to be precisely
as minimum bias \pp.  In particular, the extracted underlying
mono distributions for multiplicity and charged hadron spectra are
different from minimum bias \pp, especially in the low-\pT~region.

From this analysis, if we take the Blast-wave model at face value,
it appears as though the kinetic freeze-out of the mono collisions, around the
surface, occurs earlier than the multi, as implied by the different temperatures
at freeze-out.  This, in turn, suggests that the kinetic freeze-out
occurs at a later time such that elastic collisions still occur in the
multi after the complete freeze-out of the mono.  The radial
velocity, $\beta$, is found to be far greater for the multi, possibly
indicating a more explosive expansion.  Thus, one could argue that the
multi system is simply more densely compact.  We note that, the
Blast-wave fits are simply reflecting the difference between the
underlying mono and multi spectral shapes.

In the evolution of the collision, a chemical freeze-out occurs prior
to the kinetic freeze-out.  Chemical freeze-out occurs once all inelastic
collisions have ceased; freezing the particle species.  It is found that
this temperature ($T_{ch}$) is invariant with
centrality~\cite{cite:STAR_CuCu_FreezeOut}, thus both the mono and
multi must have the same apparent chemical freeze-out temperature.

\begin{table*}[!th]
\caption{\label{tbl:FreezeOut} Mono and multi freeze-out parameters
as derived from the least-$\chi^{2}$ fit to the mid-rapidity data from
STAR.}
\begin{tabular}{| c || c | c || c | c || c | c || c | c |}
\hline
       & \multicolumn{4}{c ||}{Au+Au} & \multicolumn{4}{ c |}{Cu+Cu} \\
Energy & \multicolumn{2}{c ||}{mono} & \multicolumn{2}{ c ||}{multi} & \multicolumn{2}{c ||}{mono} & \multicolumn{2}{ c |}{multi} \\
(GeV)  & T$_{\rm kin}$ [GeV] & $\beta$ & $T_{\rm kin}$ [GeV] & $\beta$ & $T_{\rm kin}$ [GeV] & $\beta$ & $T_{\rm kin}$ [GeV] & $\beta$ \\\hline
  200  & 0.160 & 0.163 & 0.084 & 0.639 & 0.152 & 0.268 & 0.112 & 0.555  \\
 62.4  & 0.168 & 0.184 & 0.096 & 0.576 & 0.156 & 0.226 & 0.114 & 0.548  \\ \hline
\end{tabular}
\end{table*}

\section{Summary}
In summary, we have shown that some aspects of heavy ion data can be 
reproduced using a simple two-component model, with participant-like
scaling variables.  As an initial test of the model, we find that the
multiplicity and unidentified charged hadron spectra can be reproduced
with an underlying gluon-dominated distribution; the latter without
the need for a large suppression at high-\pT.  By fitting the data, we
can simultaneously extract underlying distributions for each sub-component
of the collision: the singly-hit mono participants and the multiply-hit
multi.  From this, a suppression is observed, however, it is limited to
particles originating at surface-collisions (mono) and mostly in the low-
to intermediate-\pT~region.  The component characterizing the multiple
nucleon-nucleon interactions is found to be peaked in the intermediate
\pT~region and becomes flat at the highest measured \pT.  With these
two sub-components, we observe that there is a mechanism by which one
can reliably reproduce the multiplicity, spectral shapes, and freeze-out
parameters.  These is no need for a large (up to a factor of
five) suppression which is currently used to describe the heavy ion data
using the \ncoll~variable.
At the center of this model picture is a dense (perhaps gluon dominated)
system (multi) which is opaque to the particles produced at the periphery
of the collision (mono).  A gluon-dominated picture is not excluded by
other measurements, as discussed, the anomalous increase of
baryons, with respect to mesons, could be expected from a system with a
higher number of gluon-jets than quark-jets, though more details are
outside the scope of the current paper.

To conclude, in our model, we assert that the yield of high-\pT~particles
is not suppressed.  The apparent reduction in yields observed at
high-\pT~is in fact due to a different particle production mechanism than
what is expected from minimum bias \pp~interactions.  In particular this
mechanism does not produce high-\pT~particles at the expected binary
collision rate, but rather at a slower rate as determined by the
geometry of the collision, in a similar way to that expected from
close-packed gluons.

\begin{acknowledgments}
This work was partially supported by US DOE Grants
DE-FG02-04ER41325, 
DE-FG03-86ER40271, 
and DE-FG02-94ER40865.
\end{acknowledgments}

\end{document}